\documentclass[%
superscriptaddress,
showpacs,
 amsmath,amssymb,
]{revtex4-1}
\usepackage[colorlinks,citecolor=blue,linkcolor=blue,urlcolor=blue,bookmarks=true,hypertexnames=true]{hyperref} 
\usepackage{graphicx}
\usepackage{color}
\usepackage{dcolumn}
\usepackage{tabularx}
\usepackage{siunitx}
\usepackage{upgreek}

\usepackage[T1]{fontenc}
\usepackage{braket}

\usepackage{braket}
\usepackage{physics}
\newcommand{\be}{\begin{equation}}
\newcommand{\ee}{\end{equation}}
\newcommand{\bea}{\begin{eqnarray}}
\newcommand{\eea}{\end{eqnarray}}

\newcommand*{\hham}{\hat{\mathcal{H}}}

\begin{document}
\title{Spin-mechanics with nitrogen-vacancy centers and trapped particles}

\author{Maxime Perdriat}
\author{Clément Pellet-Mary}
\author{Paul Huillery}
\affiliation{Laboratoire De Physique de l'\'Ecole Normale Sup\'erieure, \'Ecole Normale Sup\'erieure, PSL Research University, CNRS, Sorbonne Universit\'e, Universit\'e de Paris , 24 rue Lhomond, 75231 Paris Cedex 05, France}
\author{Loïc Rondin}
\affiliation{Universit\'e Paris-Saclay, CNRS, ENS Paris-Saclay, CentraleSup\'elec, LuMIn, 91190, Gif-sur-Yvette, France}
\author{Gabriel Hétet}
\email{gabriel.hetet@ens.fr}
\affiliation{Laboratoire De Physique de l'\'Ecole Normale Sup\'erieure, \'Ecole Normale Sup\'erieure, PSL Research University, CNRS, Sorbonne Universit\'e, Universit\'e de Paris , 24 rue Lhomond, 75231 Paris Cedex 05, France}

\begin{abstract}
Controlling the~motion of macroscopic oscillators in~the~quantum regime has been the~subject of intense research in recent decades.  In this direction, opto-mechanical systems, where the~motion of micro-objects is strongly coupled with laser light radiation pressure, have had tremendous success.  In particular, the~motion of levitating objects can be manipulated at the~quantum level thanks to their very high isolation from the~environment under ultra-low vacuum conditions.
To~enter the~quantum regime, schemes using single long-lived atomic spins, such as the~electronic spin of nitrogen-vacancy (NV) centers in~diamond, coupled with levitating mechanical oscillators have been proposed.  At the~single spin level, they offer the~formidable prospect of transferring the~spins' inherent quantum nature to the~oscillators, with foreseeable far-reaching implications in~quantum sensing and tests of quantum mechanics. Adding the~spin degrees of freedom to the~experimentalists' toolbox would enable access to a~very rich playground at the~crossroads between condensed matter and atomic physics.
We review recent experimental work in~the~field of spin-mechanics that employ the~interaction between trapped particles and electronic spins in~the~solid state and discuss the~challenges ahead. Our focus is on the~theoretical background close to the~current experiments, as well as on the~experimental limits, that, once overcome, will enable these systems to unleash their full~potential.
\end{abstract}

\maketitle


\section{Introduction}

The unique control offered by single quantum systems, such as atoms or ions, has enabled an~immense boost in~the~development of quantum technologies. Extending these technologies to larger masses is important both for fundamental questions on the~nature of quantum physics at larger scales, and for the~development of innovative devices such as ultra-high precision force sensors and accelerometers~\cite{Aspelmeyer}. 

Electro-, magneto- or opto- mechanically controlled levitating objects in~vacuum are fascinating in~this regard and have been at the~focus of intense recent research activity~\cite{Millen_2020}. 
This attention is supported by the~exquisite control that one can exert over the~levitated objects. These platforms indeed allow addressing of numerous degrees of freedom, easy tuning of the~trapping potential, as well as enabling free-fall experiments~\cite{Hebestreit}, as in~atomic~physics.

Initial work proposed to coupling levitated silica nano-spheres, and even viruses, to the~optical modes of a~high finesse cavities~\cite{Chang2010, Romero-Isart2010,barker}. 
The promises of this schemes have been supported by recent experiments that reported trapped particles cooled to the~quantum regime~\cite{Delic2019b, Tebbenjohanns2020, Magrini2020, Tebbenjohanns2021}.  Active development of force sensors are in~progress and tests of various models of fundamental physics have also been proposed using various platforms ~\cite{Moore2020,Xiong2020,Vinante2020b}. 
To push these developments further and to enable operate the~mechanical oscillator in~the~quantum regime, coupling the~dynamics of the~levitated system to a~single intrinsically quantum system, such as ions, atoms, or artificial atoms, has been envisioned~\cite{Rabl, Treutlein2014}.  
Towards this goal, amongst all condensed matter system, the~negatively charged nitrogen-vacancy center (NV$^-$ center for short) in~diamond centers stands out because of the~ease with which one can optically polarize and read-out its electronic spin under ambient conditions. Most of the~proposals initially designed for clamped oscillators coupled with NV$^-$ centers~\cite{Rabl, Arcizet, Kolkowitz} can then be carried over to trapped diamonds~\cite{yin} which are, in~all current experiments, operating close to room temperature. Ensembles of NV$^-$ centers coupled identically to mechanical oscillators can also exhibit magnetic phase transitions, paving the~way towards nano-scale magnetism with long-lived and controllable spins in~a~trapped particle~\cite{ma, Wei}.  

This growing research field will drive advances in~quantum metrology via 
 quantum enhanced gyroscopy and matter-wave interferometry~\cite{ma, Scala2013, Pedernales}. 
The spin--mechanical coupling also brings important advantages for quantum sensing and metrology by providing enhanced measurement sensitivity. At the~single spin level it also features additional non-linearity and/or control that could be helpful to build non classical states of motion~\cite{yin}. Finally, they could also serve as transducers between optical and RF signals via the~mechanical mode~\cite{Rabl3}.  

In this review, we describe the~state-of-the-art levitation systems that involve NV$^-$ centers and describe their specificities and limitations. The~goal is not to draw a~comparison between the~performance of existing systems. Instead, we focus on the~important results that have been accomplished and the~remaining hurdles on the~way towards operating in~the~quantum regime with these platforms. 

In Section~\ref{sec:trapped_particle}, we briefly present the~existing levitation platform and the~equations of motion describing the~different mechanical modes. In~Section~\ref{sec:active_particle}, we present a~rapid overview of the~ancillary quantum system used to interact with the~mechanical modes, namely the~nitrogen-vacancy center in~diamond. The~Hamiltonian of the~total coupled spin--mechanical system is derived in~Section~\ref{sec:ham_spin_mecha}. Section~\ref{sec:sensing_motion} reviews experiments that demonstrate  read-out of the~mechanical motion of levitated particles using the~NV spin. Section~\ref{sec:magnetic_force_torque} and Section~\ref{sec:dynamical} provide a~classical analytical treatment of spin-bistability, spin-spring and spin-cooling in~the~adiabatic limit. Finally, Section~\ref{sec:challenges} presents the~current challenges in~spin-mechanics with trapped particles.

\section{Trapping Crystals}%
\label{sec:trapped_particle}
The basic idea behind particle levitation is to hold a~particle under atmospheric conditions or in~vacuum against gravity. Here, after presenting the~classical theoretical framework for harmonic motion analysis, we discuss current
methods for trapping crystals in~vacuum.

\subsection{Center of Mass Harmonic Motion}

The center of mass dynamics of a~levitated particle along a~direction parametrized by the~coordinate $q$, can be described by a~Langevin equation. For a~stably trapped particle, one can linearize the~trapping force so that the~particle dynamics can, to a~good approximation, be described by the~equation of a~harmonic oscillator 
\begin{equation}
	m\frac{\mathrm d^2 q}{\mathrm d t^2}+m\gamma_q \frac{\mathrm d q}{\mathrm d t} +m \omega_q^2 q= F_\mathrm{L}(t)\, ,
	\label{eq:langevin}
\end{equation}
where $\gamma_q$ is the~translational damping rate due to collisions with gas molecules and $F_\mathrm{L}$ is the~Langevin fluctuating force induced by the~interaction between the~particle and the~gas molecules, $m$ is the~particle mass, and $\omega_q$ is the~trapping frequency. Rarefied gas can be described as a~Markovian thermal bath with a~white noise spectrum so that the~fluctuating force satisfies $\langle F_\mathrm{L} \rangle=0$ and  $\langle F_\mathrm{L}(t)F_\mathrm{L}(t')\rangle=2m \gamma_q k T \delta(t-t')$ at temperatures such that $kT\gg \hbar \omega_q$.

$\gamma_q$ depends on the~exact shape of the~particle~\cite{Martinetz2018} and is proportional to the~residual gas pressure at low pressures. This property makes levitating platforms attractive, since the~thermal noise can be made arbitrary small by reducing the~gas pressure inside the~vacuum chamber. 
Note that in~practice, the~dynamics of the~particle may appear more complex than a~simple harmonic oscillator at high temperatures because of instabilities induced by the~non-linearity of the~potential ~\cite{Gieseler2013}.

\subsection{Angular Confinement: The Librational Mode}%
\label{sec:lib_mode}

When considering the~angular degree of freedom, two characteristic motional behaviours can be observed: pure rotation and libration, namely oscillation of the~particle angle about a~mean angular position.
We will see that confining the~angle so that the~libration can be described by a~harmonic oscillator is also of crucial importance when dealing with NV$^-$ centers. The~condition for librational confinement is that the~total energy in~the~angular mode must not be larger than the~angular potential depth.
Typically, angular potentials are $\pi$ or $\pi/2$-periodic so that when the~standard deviation of the~angle is in~that range, the~particle angle may jump from one angular well to another. This can be due to collisions with gas molecules for instance. 

Once confined close to an~angle $\theta=0$, the~angle follows this equation of motion 
\bea
I \frac{d^2 \theta}{d t^2} + I \gamma_\theta \frac{d \theta}{d t}+I {\omega_{\theta}}^2 \theta=\tau_L(t).
\eea
$\tau_L(t)$ are Langevin fluctuation torques satisfying $\langle \tau_\mathrm{L} \rangle=0$ and $\langle \tau_\mathrm{L}(t)\tau_\mathrm{L}(t')\rangle=2\gamma_\theta I k T \delta(t-t')$.
$\gamma_\theta$ is the~damping rate for the~angular degree of freedom due to collisions with gas molecules and $I$ the~particle moment of inertia.
As for translational modes, the~damping rate for libration is proportional to
the gas pressure at low pressures and strongly depends on the~particle shape. According to the~fluctuation dissipation theorem, the~standard deviation of the~angle $\theta$ is $
\sqrt{\langle\theta^2\rangle}=\sqrt{kT/(I\omega_\theta^2)}$. 
One can thus obtain an~approximate criterion for stable angular confinement requiring that $\sqrt{\langle\theta^2\rangle}$ is bounded by $\approx \pi/10$ in~order to ensure small angular deviations. We then obtain a~condition on the~angular rigidity $K_t=I\omega_\theta^2$ such that $K_t \gtrsim 10 \times kT$. Below this value, libration is not guaranteed, and angular deviations may be too large for efficient control of the~NV electronic spin. 


\subsection{Trapping Platforms}%
\label{sec:trap_plat}

\begin{figure}[ht!]
	\includegraphics[width=\linewidth]{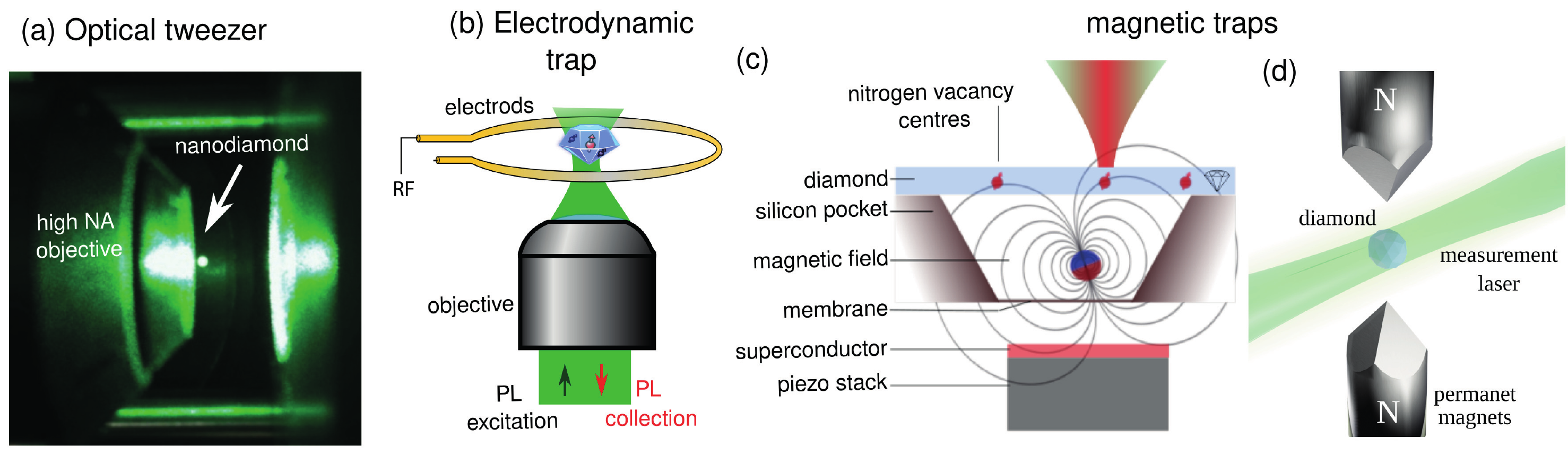}
	\caption{Different approaches to particle levitation. (\textbf{a}) Optical tweezers. Picture of an~infrared laser focused  through a~high numerical aperture (NA) objective (shown on the~left) allowing trapping a~nanodiamond at its focus. The~efficient scattering of the~green laser used for NV excitation enables one to see the~nanodiamond with the~naked eye. (\textbf{b}) Paul trap: a~charged diamond particle is held in~vacuum by electric field gradients. (\textbf{c}) A magnetic particle is levitated above a~superconductor. NV$^-$ centers implanted in~a~diamond slab above the~levitated particle can be coupled with its motion through the~magnetic field gradient. Adapted from ref~\cite{Gieseler2020a}. (\textbf{d}) A diamond particle is trapped by large magnetic field gradients thanks to its diamagnetism.}
	\label{fig:levit_setup}
\end{figure}

The original proposals for trapping particles suggested use of optical forces. Under laser illumination, small dielectric particles can indeed become polarized. The~induced dipole is then attracted to the~highest intensity region where particles are stably trapped in~three dimensions~\cite{Ashkin4853}, with typical trapping frequencies in~the~100 kHz to 1 MHz range. Reference~\cite{Jones2016} presents a~broad overview of optical tweezers. Particles can be trapped and cooled either at the~focus of a~laser beam as in~~\cite{Gieseler2012} (Figure~\ref{fig:levit_setup}a), at the~node of a~cavity field~\cite{Kiesel2013} or in~the~near-field of a~photonic crystal~\cite{Magrini2018}. Recent results in~opto-mechanics with optical tweezers and cavities are described in~a~recent review~\cite{Millen_2020}. 

Another means to trap particles, described in~Figure~\ref{fig:levit_setup}b, is to use a~Paul trap~\cite{Pau90}. The~radio-frequency modulation of a~high voltage electric field applied to the~trap electrode ensures three-dimensional confinement of charged particles with typical frequencies ranging from 100 Hz to 10 kHz. The~center of mass motion of silica nano-particles~\cite{Lorenzo}, graphene flakes~\cite{Nagornykh} and nanodiamonds~\cite{Conangla} in~Paul traps have been cooled to low temperatures under ultra-high vacuum levels using parametric feedback cooling. 

Another way to trap micro-particles is to use magnetic fields. In~magnetic traps, a~magnetic object is levitating above a~diamagnetic/superconductor material~\cite{Gieseler2020a}. Alternatively, a~diamagnetic particle---such as diamond---is levitating above magnets (see \mbox{Figure~\ref{fig:levit_setup}c).} Trapping frequencies in~the~hundreds of hertz range are typically observed. The~latter trapping approach was efficiently employed to demonstrate cooling of the~center of mass motion of trapped diamonds using feedback cooling~\cite{Hsu2015, OBrien}.

As already discussed, the~other important degree of freedom for NV spin-mechanics is the~particle angle. Trailblazing experiments with particles in~optical tweezers have shown controlled rotation of silica spheres
~\cite{Arita2013,Reimann2018,Monteiro2018, Ahn2018}. Very fast rotation frequencies up to a~few GHz have also recently been observed using dumbbell-shaped silica nanoparticles under ultra-high vacuum~\cite{Reimann2018, Ahn2018}.
When using nanodiamonds, such rotational dynamics could open the~door to observations of the~Barnett effect on long-lived electronic spins~\cite{Barnett} and to coupling spins to the~gyroscopically stabilized angular motion. 
Importantly, stable angular confinement and the~resulting librational motion of micro-particles was observed recently by several groups~\cite{VinanteA, Hoang2016, Delord2020}. 
Angular confinement can be the~result of the~shape anisotropy and birefringence of the~particles~\cite{Hoang2016}, or asymmetries of both charged particles and traps~\cite{Delord2017}.
For micro-crystals in~Paul traps, typical librational frequencies are in~the~kHz range. As discussed in~Section~\ref{sec:lib_mode}, the~moment of inertia must thus be larger than $\approx 10^{-27}$~N$\cdot$m for Brownian motion not to make trapping unstable at 300~K. Assuming a~1:2 ellipsoidal particle aspect ratio implies operation with particles with a~greater axis diameter $d\gtrsim 1~\upmu$m. In~optical traps, trapping frequencies can be as large as 1 MHz, implying that particles must have a~moment of inertia larger than $10^{-33}$~N$\cdot$m.
Particles with $d \gtrsim 50$~nm must thus be employed to harmonically confine the~angle at room temperature. The~upper bound to the~particle size will ultimately be limited by gravity, and by the~specificities of the~employed trapping mechanism. 
Observing such libration on small particles is an~important step forward, because it will facilitate torque sensing with individual spins (see~Section~\ref{sec:magnetic_force_torque}). 
 
\section{Coupling to an~Ancillary Quantum System: The Special Case of the~NV$^-$ Center}%
\label{sec:active_particle}

As discussed in~the~introduction, coupling a~levitated particle to an~ancillary quantum system is an~exiting way to extend its capabilities.
Coupling the~internal electronic state of atoms or ions to trapped mechanical oscillators is intensively studied ~\cite{Ranjit2015a, Bykov} and was successfully demonstrated with clamped oscillators ~\cite{Treutlein2014,Karg2020,Thomas2020}. 
Colloidal quantum dots~\cite{Salakhutdinov2020}, rare-earth ions in~a~solid matrix~\cite{Rahman2017}, or color centres in~semi-conductor particles~\cite{Pettit2017,Delord2017a,Hoang2016a} have all been levitated in~optical tweezers. These objects feature defects that behave as atoms which are hosted in~a~solid-state matrix, so these experiments are important steps forward.
One important extra criterium is the~coupling strength between the~ancillary systems' quantum state and the~levitated particle dynamics.
The coupling strength quantifies how much the~motion of the~object affects the~atomic state, as well as how much the~change in~the~quantum system impacts the~motion.
We will describe this coupling more quantitatively in~Section~\ref{sec:ham_spin_mecha} using the~NV$^-$ center.

\subsection{The NV$^-$ Center}

The NV$^-$ color centers in~diamond are of particular interest for spin-mechanics. First, thanks to the~intense research activities around these color centers, triggered by their potentials in~quantum information processing and the~development of innovative sensors, their internal electronic and nuclear spin properties are very well understood. For the~same reason, the~control of diamond materials has progressed significantly over the~past decade.  
Lastly, the~electronic spin of the~NV$^-$ center can easily be coupled with the~diamond host matrix's motion using a~magnetic field~\cite{Rabl, ma, Delord2017, Lee_2017, Ge, Ovartchaiyapong, Teissier}.

The physics and applications of the~NV$^-$ centers have already been discussed in~recent reviews ~\cite{Rondin2014, DOHERTY20131}. We simply recall here the~basics of NV$^-$ physics required for understanding the~coupling to trapped particles.

\begin{figure}[htb!]

\includegraphics[width=0.99\linewidth]{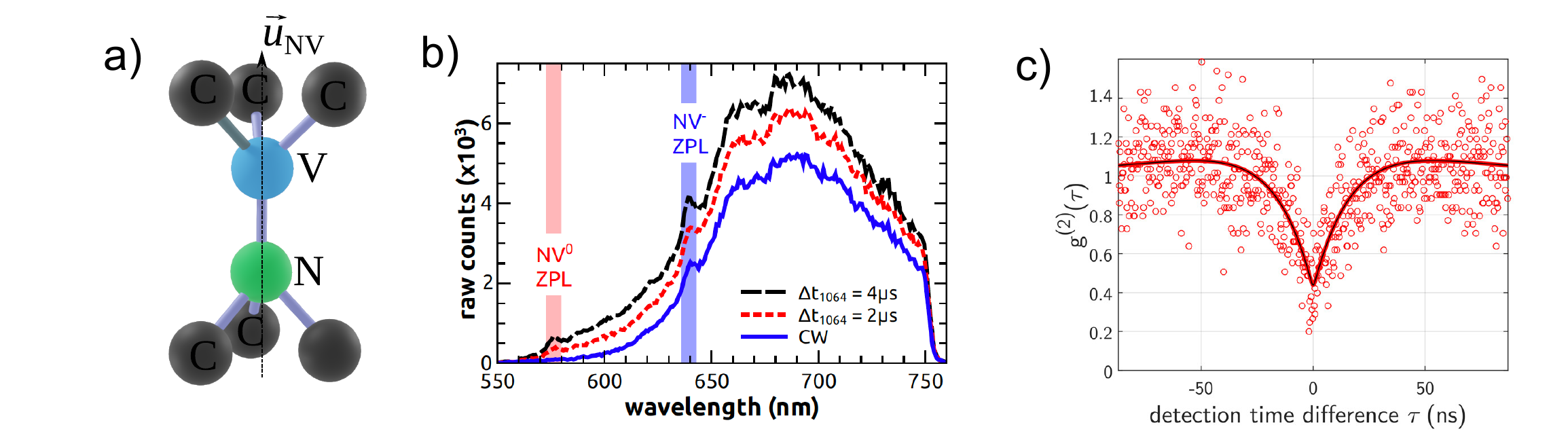}
\caption{(\textbf{a}) Schematic of the~crystalline structure of NV defect in~diamond. The~axis $N-V$ is the~quantization axis.  (\textbf{b}) Photoluminescence from NV$^-$ centers in~an optical tweezer using three different excitation conditions. Adapted with permission from~\cite{Neukirch2013}. © The Optical Society.  (\textbf{c}) Autocorrelation function of photon emission from a~single NV$^-$ center in~a~nanodiamond levitated in~a~Paul trap. Adapted with permission
from~\cite{Conangla}. Copyright 2020 American Chemical Society}
\label{CaracNV}
\end{figure}

The NV color center is a~point defect inside the~diamond matrix consisting of a~substitutional nitrogen atom (N) combined with a~vacancy (V) in~one of the~nearest neighboring sites of the~diamond crystal lattice as depicted in~Figure~\ref{CaracNV}a.
This defect behaves as an~artificial atom hosted by the~diamond matrix. It combines unique luminescence and spin properties.
Indeed, because its energy levels are well within the~large band gap of diamond, the~NV center has an~extremely stable luminescence in~the~near-infrared. 
It can be found in~two different charge states: the~NV$^-$ and NV$^0$.
The NV$^-$ zero-phonon line is at around $\lambda_\mathrm{ZPL}=637$~nm. It is associated with broad phonon sidebands that extend up to $\approx$750 nm. The~NV$^-$ is the~most interesting for spin--mechanical interactions as we will see.

The NV$^-$ luminescence can readily be accessed using standard confocal microscopy, and can straightforwardly be adapted for levitation platforms. The~photoluminescence from NV$^-$ centers in~trapped diamonds under atmospheric conditions or in~vacuum was observed by several groups~\cite{Kuhlicke, Neukirch2015, Pettit2017, Neukirch2013, Delord2017a, Hsu}. 
Figure \ref{CaracNV}b shows the~photoluminescence from NV$^-$ centers in~optical tweezers observed in~\cite{Neukirch2013}. Different pulsing sequences were used in~order to mitigate PL quenching from the~1064 nm trapping laser. Note that, alternatively, photoluminescence (PL) quenching can be greatly reduced using other trapping laser wavelengths~\cite{Hoang2016a}.
Further studies demonstrated trapping of nanodiamonds containing single NV$^-$ centers~\cite{Conangla, Neukirch2015,Pettit2017}.
 Figure \ref{CaracNV}c shows the~autocorrelation function of the~PL from a~single NV$^-$ center inside a~nanodiamond trapped in~a~Paul trap~\cite{Conangla}. The~antibunching dip at zero delay, below 0.5 is a~proof for the~presence of a~single NV$^-$ center inside the~trapped diamond, which opens a~path towards non-gaussian spin-mechanics using the~NV$^-$ spin.

\subsection{The NV$^-$ Center Electronic Spin}

\begin{figure}[htb!]
	
	\includegraphics[width=0.98\linewidth]{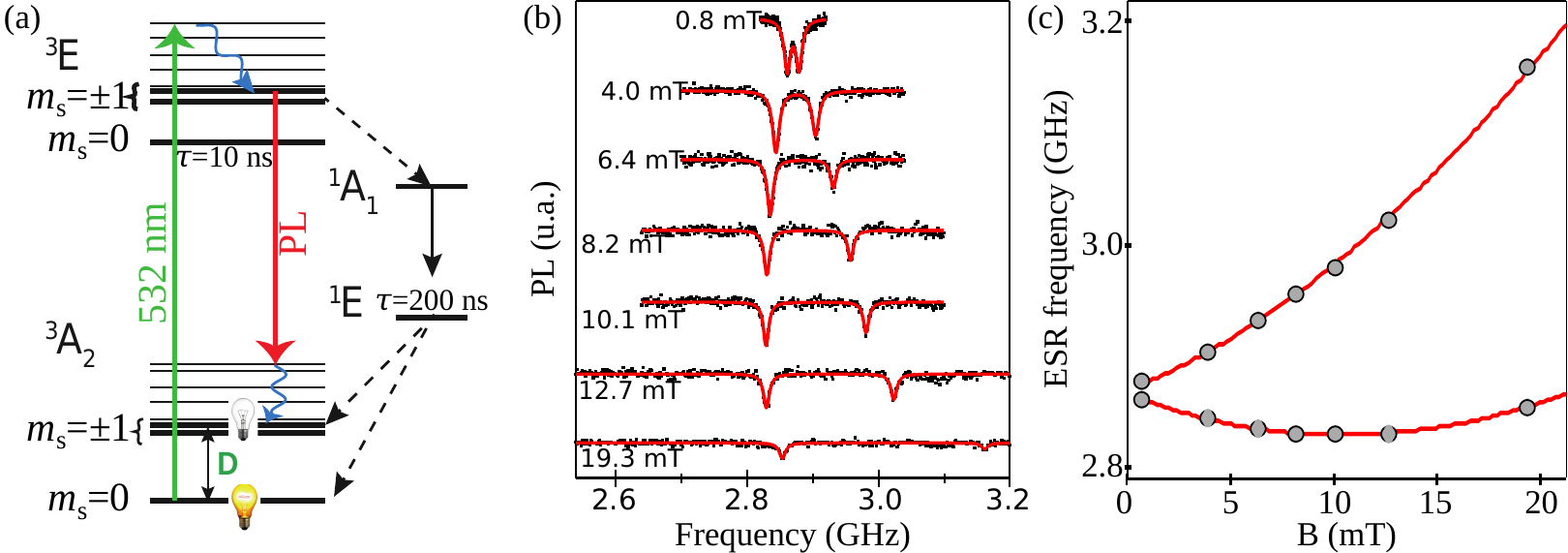}
    \caption{NV$^-$ center level structure. (\textbf{a}) Simplified electronic structure of the~NV$^-$ center. (\textbf{b})~Example of optically detected magnetic resonance on a~single NV$^-$ center in~a~bulk diamond. The~magnetic field is applied with an~angle  $\theta=74^\circ$ with respect to the~NV axis. (\textbf{c}) ESR frequencies from (\textbf{b}) (gray dots) compared with the~full theory (red lines) obtained by computing the~NV Hamiltonian eigenvalues (Equation~\eqref{eq:NVHam}). Figure (\textbf{b},\textbf{c}) are adapted from~\cite{Tetienne2012}.}%
	\label{PLNV}
\end{figure}

In addition to the~possibility of observing stable PL at ambient conditions, the~NV$^-$ center also carries a~spin that can be manipulated at room temperature.
The electronic spin of the~NV$^-$ center in~the~ground state is a~spin triplet $S=1$, with a~quantization axis $\textbf u_\mathrm{NV}$ enforced by the~crystalline field around the~defect. The~spin projections along this axis are labelled with the~quantum number $m_s$. Due to a~spin--spin interaction between the~two electrons in~the~ground state the~$\ket{m_s=\pm 1}$ states are split from the~$\ket{m_s=0}$ by the~zero-field splitting $D\approx 2.88$~GHz (see Figure~\ref{PLNV}a). 
These states are purely magnetic states, implying a~very long longitudinal decay time $T_1$, on the~order of milliseconds in~typical diamond samples~\cite{Jarmola_temperature_2012}. 
Note that in~strained diamond, or in~the~presence of local electric fields~\cite{Mittiga}, the~states $\ket{m_s=\pm 1}$ can also be split, by a~parameter often denoted $2E$. This splitting is typically a~few MHz in~nanodiamonds. In~practice, a~magnetic field bias larger than $E$ is employed in~order to reach large spin--mechanical couplings, so we neglect this zero-field splitting in~the~following.

One interesting aspect of the~NV$^-$ center resides in~the~possibility to optically initialize and read-out its electronic spin state. Indeed, under green laser irradiation the~NV$^-$ center is polarized in~the~$\ket{m_s=0}$ state, while the~photoluminescence level depends on the~populated spin state. The~$\ket{m_s=0}$ state can be up to 30~\% brighter than the~$\ket{m_s=\pm 1}$ states~\cite{DOHERTY20131}.
When a~microwave tone is resonant with one of the~transitions from $\ket{m_s=0}$ to $\ket{m_s=\pm 1}$, a~drop of luminescence can then observed. This forms the~basis of \emph{Optically Detected Magnetic Resonance} (ODMR). 
The energy of these transitions can be found from the~Hamiltonian of the~electronic spin of the~NV$^-$ center in~the~ground state, which reads
\begin{equation}
	\hham/\hbar = D \hat{S}_z^2 +\gamma_e \hat{\textbf{S}} \cdot \textbf{B},
    \label{eq:NVHam}
\end{equation}
where $\gamma_e$ is the~gyromagnetic factor of the~electron.
Because of the~zero-field splitting $D$ and the~spin 1 character of the~NV$^-$ center, the~eigenstates of the~Hamiltonian depend upon the~angle between the~magnetic field and the~NV$^-$ center when $\gamma_e B<D$.

Typical single NV$^-$ ODMR spectra with fixed diamonds are presented in~Figure~\ref{PLNV}b showing the~transitions from the~$\vert 0\rangle$ to $\vert \pm 1\rangle$ states as a~drop of the~PL. For this experiment, the~NV$^-$ is at an~angle of 74$^\circ$ with respect to the~magnetic field. 
Figure~\ref{PLNV}c shows the~change in~the~frequency of the~two transitions as a~fonction of magnetic field as well as a~numerical calculation using Equation~(\ref{eq:NVHam}).

The width of the~observed electron spin resonance (ESR) dip is here limited by the~dipolar coupling between the~spin of NV$^-$ centers to other paramagnetic impurities in~the~diamond. The~most dominant impurity is the~substitutional nitrogen (P1 centers), with concentrations ranging from 100 to 500 ppm in~high-pressure-high-temperature grown diamonds (see Section~\ref{sec:production}). Dipolar coupling between the~NV$^-$ center and P1 centers typically lead to an~inhomogeneous dephasing time $\Gamma_2^*$ on the~order of 5 to 10 MHz~\cite{Bauch}. 
Ultimately, the~linewidth of the~ESR is of crucial importance, since it limits the~measurement sensitivity as well as impacting the~spin coupling strength to the~mechanics. Nevertheless, as discussed in~the~Sec.~\ref{sec:dynamical}, diamond growth allows fine control over impurities. 

The frequency shifts of the~spin resonances with magnetic field and their dependence with the~magnetic field angle are important ingredients for coupling the~center of mass and the~angle of levitated particles to the~NV$^-$ spin states, as we now discuss.

\section{Hamiltonian of the~Spin--Mechanical System}%
\label{sec:ham_spin_mecha}
In the~following, we consider a~diamond containing a~single NV$^-$ center in~the~presence of a~homogeneous magnetic field as well as in~a~magnetic field gradient. 
We will consider the~case of an~NV$^-$ center in~a~trapped diamond but this analysis can be carried over to studies where an~NV$^-$ center is coupled with a~distant trapped magnet~\cite{Huillery2020, Gieseler2020a}. 

When the~genuine quantum nature of single spins is not of interest, the~following treatment can also straightforwardly be adapted to $N$ spins. If correlations between the~NV spins inside the~diamond from direct dipolar interactions~\cite{Bachelard,PANAT,Venkatesh,Juan}, or resonator mediated interactions~\cite{ma, Wei, Schuetz} are neglected, the~coupling strength can simply be multiplied by the~number of spins. 
To be able to directly carry a~single-spin analysis over to the~ensemble case, one must also assume that the~NV centers all have the~same orientation in~the~diamond. In~practice this is not the~case since the~NV$^-$ centers will typically be found with the~same probability along the~four $[111]$ 
 diamond directions. 
However, in~the~presence of a~magnetic field that does not broaden the~ESR width, the~microwave can select only one of the~eight resulting ESR transitions so that one can treat this problem using $N$ effectively spin 1/2 systems in~a~single orientation with ESR frequencies that are within the~inhomogeneous broadening $\Gamma_2^*$.

We will study the~coupling between the~spin and the~two main degrees of freedom of a~levitating mechanical oscillator: the~center of mass motion (CoM) and the~libration.
We will assume that there is no coupling between these two degrees of freedom so that we can treat them separately. Note that this may not be true generally. For instance, in~Paul traps, if the~charge distribution has a~non-zero dipole component, the~center of mass and the~angle may become coupled~\cite{Martinetz20}. 
Similarly, shape anisotropy or birefringence of optically trapped particles can also induce a
mode coupling~\cite{Arita2020SA,Trojek2012JOSAAJ}. These mechanisms are neglected here, but can be of interest for transferring the~quantum state of one spin-coupled mode to another one.

\subsection{Coupling to the~Center of Mass}

The center of mass motion (CoM) of the~trapped particle can be coupled with the~NV$^-$ spin using a~magnetic field gradient~\cite{Rabl}. 
Let us reduce the~study of the~spin-CoM coupling to a~1D problem. 
We assume that the~NV axis is along the~magnetic field. This greatly simplifies the~problem by leaving aside mixing between the~NV eigenstates and coupling between the~CoM and the~libration. When the~magnetic field gradient is in~the~$z$ direction: $\textbf{B} = B_0\textbf{e}_z+ \frac{\partial B_z}{\partial z }z\textbf{e}_z$ to first order in~the~position. Under those assumptions, the~Hamiltonian of the~spin--mechanical system reads $\hham_{\rm com}= \hham_{\rm mecha}+ \hham_{\rm NV}$, where $\hham_{\rm mecha}= \frac{p_{z}^2}{2m}+\frac{1}{2}m \omega_{z}^2 z^2 $ and $\hham_{\rm NV}=\hbar D  \hat{S}_{z}^2  +\hbar  \gamma_e( B_0 + \frac{\partial B_z}{\partial z } z)  \hat{S}_z $ .

Due to the~Zeeman term in~$\hham_{\rm NV}$, the~magnetic energy is linearly proportional to the~particle position. 
A magnetic force directly related to the~magnetic field gradient and the~chosen spin polarisation can thus be applied to the~particle (see Section~\ref{sec:magnetic_force_torque}).

In the~presence of an~oscillating magnetic field $B_{\mu W}$ that is linearly polarized along the~$x$ direction, the~spin part of the~Hamiltonian reads 
\begin{align}
\hham_{{\rm NV}+\mu\rm w}=  \hbar D \hat{S}_{z}^2 + \hbar \gamma_e(B_0  +  \frac{\partial B_z}{\partial z } z )  \hat{S}_z + \hbar \Omega \cos({\omega t}) \hat{S}_{x}.
\end{align}
 where $\Omega= \gamma_e B_{\mu W}$.
Moving to the~rotating frame at the~microwave frequency $\omega$ through the~unitary transform $\hat U=e^{i\omega t \hat{S}_{z}^2}$ and making a~rotating wave approximation, we get \begin{align}
\hham_{{\rm NV}+\mu\rm w}'=  \hbar  (D-\omega) \hat{S}_{z}^2+ \hbar  \gamma_e (B_0  +  \frac{\partial B_z}{\partial z } z ) \hat{S}_z + \hbar  \frac{\Omega}{2}  \hat{S}_{x}.
\end{align}

If $\gamma_e B_0$ is greater than both $2E$ and $\Gamma_2^*$ and if the~optical pumping process is stronger than the~relaxation rate $T_1$ (see Section~\ref{sec:active_particle}), resonantly tuning the~microwave frequency to the~$\ket{0} \leftrightarrow \ket{+1} (resp. \ket{-1})$ transition enables the~$\ket{-1}$ (resp. $\ket{+1}$) state to be safely neglected. Choosing a~microwave signal tuned to the~transition $\ket{0} \leftrightarrow \ket{+1}$, we finally obtain the~Hamiltonian of a~two-level system coupled with a~mechanical oscillator:
\begin{align}
\hham_{\rm total}=  \hham_{\rm mecha}+ \hham_{\rm Spin}+\hham_{\rm Spin-mecha},
\label{eq:ham_total}
\end{align}
where $\hham_{\rm Spin}=-\hbar\Delta\ket{1}\bra{1}+ \hbar\frac{\Omega}{2} (\ket{0}\bra{1}+\ket{1}\bra{0})$  with the~microwave detuning $\Delta=\omega-D-\gamma_e B_0$
and $\hham_{\rm Spin-mecha}=\hbar G_z z  \ket{1}\bra{1}$ with the~coupling constant $G_z=\gamma_e  \frac{\partial B_z}{\partial z }$. Note that the~latter has been redefined a~$\Omega\equiv\frac{\Omega} {\sqrt{2}}$ to ensure normalisation of the~$\hat{S}_x$ operator.

Before studying this Hamiltonian and the~related experiments in~more details, we will discuss the~spin-coupling to the~librational degree of freedom.
\subsection{Coupling to the~Libration}

The theory behind spin-coupling to the~libration has been presented in~several papers~\cite{Ge, ma, Delord2017}.  
Here, we derive a~simplified Hamiltonian in~the~$\gamma_e B\ll D$ limit. 
Let us use the~vectors $(0,\textbf{e}_x,\textbf{e}_y,\textbf{e}_z)$ to specify the~orientation of the~laboratory frame and the~vectors $(0,\textbf{e}_{x'},\textbf{e}_{y'},\textbf{e}_{z'})$ to specify the~particle frame. We choose $\textbf e_z'$ as the~anisotropy axis of the~NV$^-$ center in~the~crystalline structure of the~diamond. The~three Euler angles operators $({\hat{\phi}},{\hat{\theta}},{\hat{\psi}})$ describing the~angular position of the~diamond are chosen in~the~$(zy'z'')$ convention. The~magnetic field is supposed to be homogeneous and its direction is fixed in~the~laboratory frame. It is chosen along the~$z$ direction, so that $\textbf{B}=B \textbf{e}_z$.

The Hamiltonian of the~spin--mechanical system in~the~laboratory frame reads
\begin{align}
\hham_{\rm lib}= \frac{{\hat{\textbf{L}}}^2}{2I} +U(\hat{\phi},\hat{\theta},\hat{\psi}) + \hbar D \hat{S}_{z'}^2 + \hbar \gamma_e B \hat{S}_z,
\end{align}
where ${\hat{\textbf{L}}}$ is the~angular momentum operator of the~diamond in~the~laboratory frame, $U(\hat{\phi},\hat{\theta},\hat{\psi})$ is the~angular confining potential and $\hat{S}_z$,$\hat{S}_{z'}$ are NV$^-$ center spin operators.
Contrary to when studying the~coupling of the~NV center to the~center of mass, the~NV direction is not  necessarily fixed in~the~laboratory frame. This implies that the~$\hat{S}_{z'}$ operator depends on angular operators $({\hat{\phi}},{\hat{\theta}},{\hat{\psi}})$ which do not commute with the~diamond angular momentum operator $\hat{\textbf{L}}$. Consequently, the~commutator $\commutator {\hat{\textbf{L}}^2}{\hat{S}_{z'}^2}\neq 0$~\cite{Rusconi}. 

In the~following, we will restrict the~study to one librational mode that is assumed to be in~the~$(zz')$ plane formed by the~magnetic field and spin direction. The~diamond angular motion is parametrized by the~nutation angle operator $\hat{\theta}$. $\theta'$ represents the~equilibrium angular position of the~diamond.

The Hamiltonian of the~simplified system reads
\begin{align}
\hham_{\rm lib}= \frac{{\hat{p}}_{\theta}^2}{2I} + \frac{1}{2}I \omega_\theta^2 (\hat{\theta}-\theta')^2+  \hbar D \hat{S}_{z'}^2 + \hbar \gamma_e B \hat{S}_{z}, 
\end{align}
where ${\hat{p}}_{\theta}$ is the~angular momentum operator along the~$y$ axis.
Moving to the~particle frame through the~unitary transformation $\hat{U}=e^{i{\hat{\theta}} \hat{S}_y}$ changes the~Hamiltonian to
\begin{align}
\hham'_{\rm lib}= \frac{({\hat{p}}_{\theta}-\hbar \hat S_y)^2}{2I} +\frac{1}{2}I \omega_\theta^2 (\hat{\theta}-\theta')^2+  \hbar D \hat{S}_{z}^2 + \hbar \gamma_e B\left(\cos{{\hat{\theta}}} \hat{S}_{z}-\sin{{\hat{\theta}}} \hat{S}_{x}\right).
\end{align}
In this frame, ${\hat{p}}_{\theta}$ is changed to ${\hat{p}}_{\theta}-\hbar \hat S_y$, which is the~total angular momentum of the~spin--mechanical system along the~$y$ axis. It is the~sum of the~particle and the~NV-spin angular momentum. 
This modified angular momentum could cause precession of the~particle. Such a~precession is predicted to be observable in~levitated hard ferromagnetic nano-particles~\cite{Kimball, Rusconi3}, with important applications in~gyroscopy.
For nano- or micro- particles containing a~smaller amount of spins, the~total angular momenta correction $\hbar N \hat{S}_y$ is negligible.
In order to estimate the~relevance of this term in~current experiments, one can note that $\left \langle{\hat{p}}_\theta \right \rangle\approx \sqrt{k T I}$ is far above $\hbar N$ even when using a~highly-doped ($\approx$10 ppm) micron-sized diamond at temperatures above $\mu$K. 
This condition will thus be verified for all smaller particles, since the~maximal density of NV centres scales with the~particle volume.
In this considered temperature regime, we can thus treat the~angle as a~scalar and neglect the~contribution from the~angular spin momentum, bearing in~mind that this approximation drops when using an ultra-cold oscillator.

The remaining step is to diagonalize the~spin part of the~Hamiltonian. Apart from in~ref.~\cite{Perdriat}, experiments currently operate in~the~regime $\gamma_e B \ll D$. We thus assume this condition to be fulfilled in~the~following. We also assume small mixing between the~two NV$^-$ excited states, which is satisfied under the~condition $\sin^2{\theta'} \ll \cos{\theta'}$. Lastly, we change variable and shift the~angle ${\theta}$ to the~equilibrium position $\theta'$. Under those assumptions, the~Hamiltonian reads (see Appendix~\ref{Appendix A} or the~detailed calculation):
\begin{align}
\hham'''_{\rm lib} \simeq \frac{{{p}}_{\theta}^2}{2I} + \frac{1}{2}I \omega_\theta^2 {\theta}^2+ \hbar (\omega_{+1}({\theta}) \ket{+1'}\bra{+1'} + \omega_{0}({\theta}) \ket{0'}\bra{0'}+ \omega_{-1}({\theta}) \ket{-1'}\bra{-1'}).
\label{eq:Ham_lib'''}
\end{align}

\begin{figure}[htbp!]
	
	\includegraphics[width=0.90\linewidth]{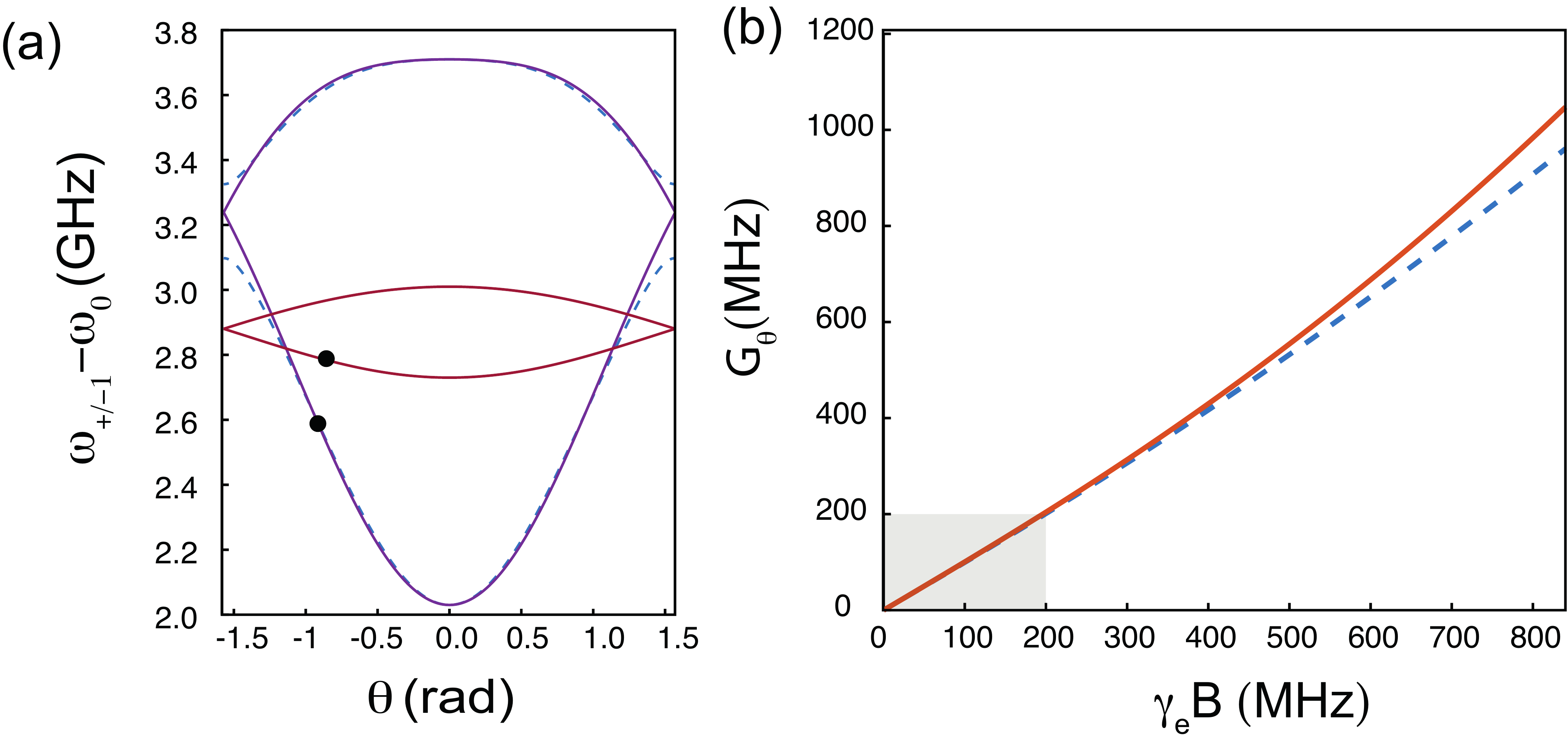}
	\caption{{(\textbf{a}) Electronic} spin transition energies of the~NV$^-$ center as a~function of the~angle $\theta$ for magnetic fields of 50 G (brown lines) and 300 G (blue lines). The~black dots indicate the~optimum angles for the~largest spin-torques. b)
	Optimum coupling strength $G_\theta$ on the~$\ket{0}$ to $\ket{-1}$ transition as a~function of the~external magnetic field. The~shaded area corresponds to the~magnetic field values for which the~optimized coupling strength is very close to $\gamma_e B$.  In both plots, numerical calculations without approximations are shown by dashed lines.
	}%
	\label{angleopt}
\end{figure}

The expression of the~new eigenstates $\ket{\pm 1'}$ and $\ket{0'} $ is listed in~Appendix~\ref{Appendix A}.
The frequencies $\omega_{i}({\theta})=\omega_{i}+\beta_{i}{\theta}$ of the~spin resonance are plotted as a~function of the~angle $\theta$ for two different magnetic field values in~Figure~\ref{angleopt}a. 

It was shown in~\cite{Tetienne2012} that the~pumping process is not modified to first order in~the~transverse magnetic field. We can thus neglect the~mixing contribution from the~optical pumping process of the~NV$^-$ center. This is also manifest in~Figure \ref{PLNV}b where the~contrast of the~ODMR drops significantly only with magnetic field values above $\approx$10~mT. 

Similar to the~calculation for the~center of mass (CoM), we obtain the~Hamiltonian (see Appendix~\ref{Appendix A}):
\begin{align}
\hham'_{{\rm NV}+\mu\rm w}/\hbar \simeq -\Delta_{+1}({\theta}) \ket{+1'}\bra{+1'} -  \Delta_{0}({\theta}) \ket{0'}\bra{0'}- \Delta_{-1}({\theta}) \ket{-1'}\bra{-1'}+ \frac{\Omega}{2} \hat{S}_{x},
\end{align}
where $\Delta_{+1}({\theta})=\omega-\omega_{+1}({\theta})$, $\Delta_{-1}({\theta})=\omega-\omega_{-1}({\theta})$ and $\Delta_{0}({\theta})=-\omega_{0}({\theta})$. We redefined $\Omega\equiv \Omega \cos{\theta'}$.
The eigenstates are listed in~Appendix A.
Choosing a~microwave tuned close to resonance with the~$\ket{0'}$ to $\ket{+1'}$ transition allows us to restrict the~study to the~Hamiltonian of a~two-level system 
\begin{align}
\hham''_{{\rm NV}+\mu\rm w}/\hbar \simeq  \left(-\Delta+\beta_1 {\theta}\right) \ket{1'}\bra{1'}+ \beta_0 {\theta} \ket{0'}\bra{0'} + \frac{\Omega}{2}(\ket{0}\bra{1'}+\ket{1'}\bra{0}).
\end{align}

Here $\Delta=\Delta_{+1}-\Delta_0$ and $\Omega\equiv\frac{\Omega} {\sqrt{2}}$ to take into account the~$\sqrt{2}$ factor in~the~$\hat{S}_x$ operator of the~spin-1 system. We have fixed the~energy reference to be the~energy of the~$\ket{0'}$ state.

The second term $ \beta_0 {\theta} \ket{0'}\bra{0'}$ operates as a~small shift of the~angular position when the~NV$^-$ ground state is populated and can be interpreted as a~consequence of van Vleck paramagnetism. The~mixing between the~ground and excited states induced by the~transverse magnetic field indeed generates a~non-zero magnetic moment in~the~NV$^-$ center~\cite{pellet2021magnetic, Perdriat}. 
This term gives rise to a~new equilibrium position in~experiments because the~$\ket{0'}$ is populated by the~green laser. 

At this point, a~word of caution is in~order if one wishes to translate the~above description to ensembles of NV centers.
We indeed need to check for the~consistency of our initial assumptions about independent NV directions in~the~presence of strong van Vleck paramagnetism. This paramagnetism is present in~the~absence of microwave excitation, so all NV orientations contribute to give a~spin-torque in~the~ground state, the~extent of which depends upon the~transverse magnetic field amplitude for each orientation.
When including the~four orientations, the~first-order effect is a~slight reduction of the~van Vleck paramagnetic susceptibility that can be estimated for a~single orientation. This is due to an~overall spatial averaging of the~magnetizations from the~four NV orientations, which should be recast in~$\beta_0$. 
The reader can find more information in~\cite{pellet2021magnetic}.

Let us redefine the~center of the~mechanical resonator to be around the~equilibrium angular position when the~population is in~the~$\ket{0'}$ state 
by including the~van Vleck shift in~$\Delta$.  Then, by defining $G_\theta=\beta_1-\beta_0$ to be the~single-spin mechanical constant, the~Hamiltonian of the~system can finally be written as the~simple two-level atom Hamiltonian:
\bea
\hham''_{{\rm NV}+\mu\rm w}/\hbar= \begin{pmatrix} -\Delta(\theta)  & \Omega/2  \\ \Omega /2 & 0 \end{pmatrix},
\label{TLSHam}
\eea
where $\Delta(\theta)=\Delta-G_\theta{{\theta}}$ is the~frequency difference between the~states $\ket{0'}$ and $\ket{1'}$. 
The Hamiltonian for the~librational mode is thus the~same as for the~center of mass (\mbox{Equation~\eqref{eq:ham_total}}), when replacing the~coupling constant $G_z$ by $G_\theta$.

The optimum coupling strength $G_\theta$ on the~$\ket{0}$ to $\ket{-1}$ transition is plotted in~Figure~\ref{angleopt}b. Ones finds that $G_\theta\approx \gamma_e B$ when $\gamma_e B< 200$~MHz. There is a~good match between analytical numerical calculations in~this range of magnetic fields. Slight deviations between analytical treatment and numerical calculations are visible when $\gamma_e B> 200$~MHz due to stronger state mixing as $\gamma_e B$ approaches $D$. 

\section{Sensing the~Motion of a~Trapped Particle Using NV$^-$ Centers}
\label{sec:sensing_motion}

As we just described, the~ESR frequencies depend strongly on the~angle of the~homogeneous magnetic field with respect to the~NV axis, and on the~center of mass in~a~magnetic field gradient. 
In the~presence of a~microwave drive on the~diamond, the~change in~the~photoluminescence from the~NV$^-$ centers can then be used as a~marker of the~center of mass motion of diamonds or distant ferromagnetic particles, as pioneered in~experiments with tethered oscillators~\cite{Arcizet, Kolkowitz}.
With trapped diamond particles, the~angular dependence of the~NV center energy levels also means that it is possible to detect their full rotation and even their librational motion. 

\begin{figure}[htb!]

\includegraphics[width=0.99\linewidth]{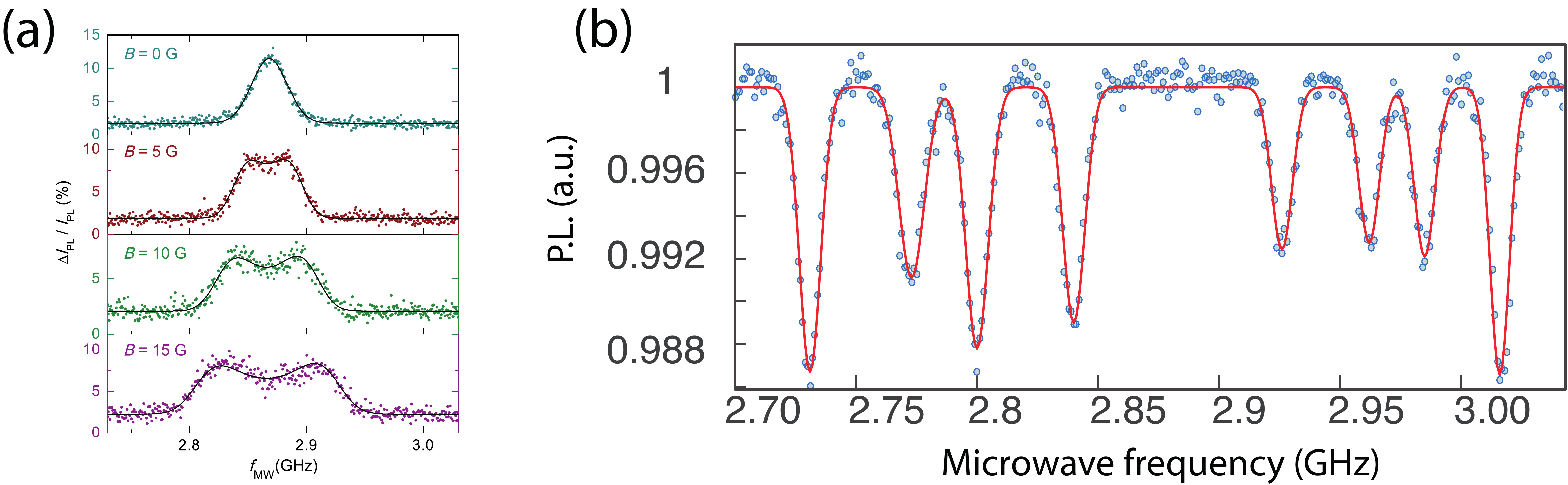}
\caption{ (\textbf{a}) ODMR for small ensembles of nano-diamond trapped in~water, under increasing magnetic fields. Adapted form~\cite{Horowitz2012}. (\textbf{b}) ODMR spectrum for a~small ensemble of NV$^-$ centers in~a~levitated micro-diamond in~a~magnetic field of about 50 G. Eight peaks corresponding to the~four orientations of the~NV$^-$ centers, are observed. Reprinted figure with permission from~\cite{Delord2018}. Copyright 2018 by the~American Physical Society.}%
\label{ODMR}
\end{figure}

Figure~\ref{ODMR}a shows an~ODMR obtained from optically trapped diamonds in~water featuring broad lines due to significant angular Brownian motion in~the~presence of a~magnetic field~\cite{Horowitz2012}. 
Similar ODMR shapes were observed in~\cite{Delord2017a, Neukirch2015}.
These results show that NV centers are an~efficient tool to measure the~rotation of particles. Further, the~fully rotating regime may also shed new light on geometric phases acquired by NV centers as well as offering perspectives for  
efficient sensing of magnetic fields in~the~rotating frame~\cite{Wood}.

Note that, additionally, control of the~trapping laser polarisation allows tuning the~NV angles
~\cite{Geiselmann}.  In~\cite{Delord2017a, Delord2018}, ODMR from diamond particles in~Paul traps was also reported on angularly stable micro-diamonds (with typical librational frequencies in~the~kHz range). Figure~\ref{ODMR}b shows the~PL of NV$^-$ centers from a~diamond containing about 1000 NV$^-$ centers. Eight ODMR lines were observed due to the~four projections of the~magnetic field onto the~four possible NV axes in~the~diamond cristalline structure and thanks to the~efficient particle angular confinement. Assuming thermalization with a~gas at a~temperature close to T=300~K, the~angular standard deviation is $\Delta \theta \approx$ 1~mrad (see Section~\ref{sec:trapped_particle}).  
The expected extra broadening of the~ODMR lines in~a~magnetic field $B\approx 50$ G is on the~order of $G_\theta \Delta \theta \approx \gamma B \Delta \theta=150$ kHz. This value is smaller than the~width given by the~dipolar coupling between the~NV and P1 centers (see Section~\ref{sec:active_particle}).

Two recent experiments have even shown read out of the~harmonic motion of a~trapped particle using NV centers~\cite{Huillery2020, Gieseler2020a}. 
In the~first experiment~\cite{Huillery2020}, a~hybrid diamond-nickel particle was levitated in~a~Paul trap. The~NV$^-$ centers in~the~diamond were employed to read-out the~librational motion.
In the~second experiment~\cite{Gieseler2020a}, a~magnet was levitating on top of superconducting sheet and a~diamond containing a~single NV$^-$ center was brought in~the~vicinity of the~magnet to read-out its center of mass motion.

\begin{figure}[htbp!]
 {\includegraphics[width=0.7\textwidth]{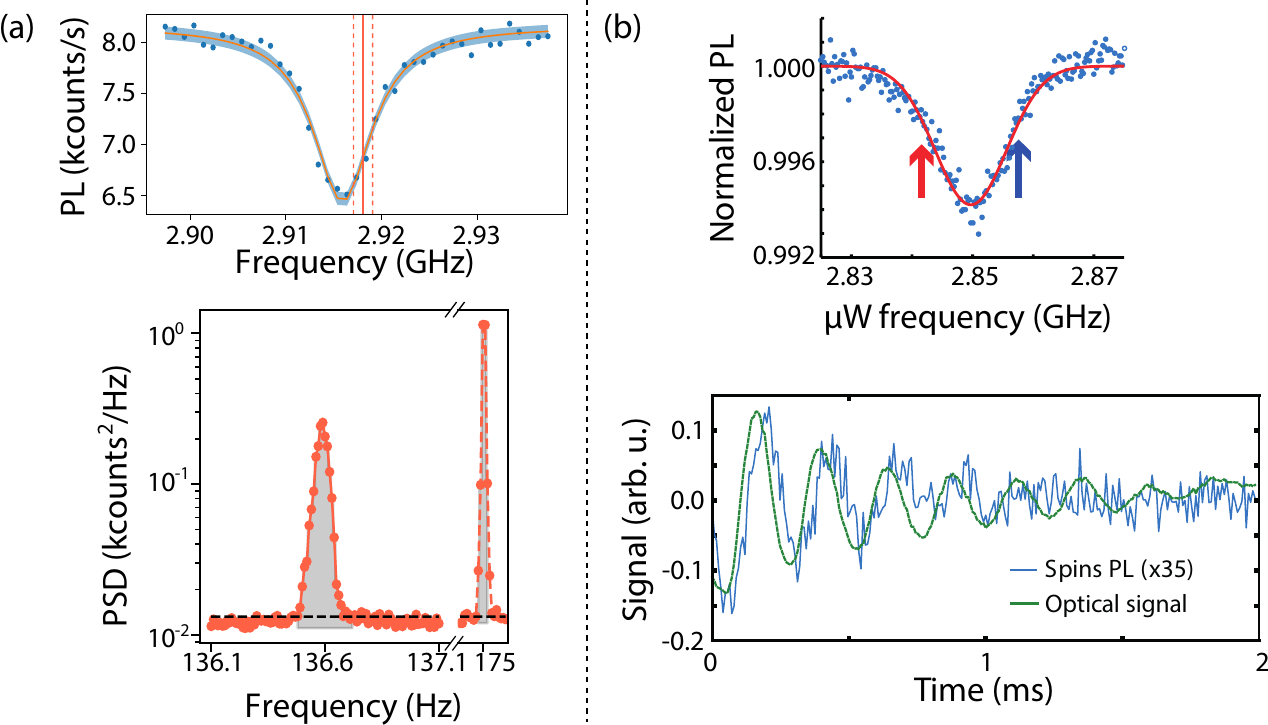}}
  \caption{NV$^-$
 sensing of the~motion of trapped particles. (\textbf{a}) Top trace: ODMR on a~$\vert m_s=0\rangle$ to $\vert m_s=+1\rangle$ transition from a~fixed single NV$^-$ center, located 100 $~\upmu$m away from the~levitating magnet. Below: PSD of the~NV fluorescence signal when a~microwave is tuned to the~blue side of the~ODMR peak. The~right narrow peak shows the~magnet motion when the~microwave is modulated at 175 Hz. Adapted from~\cite{Gieseler2020a}. (\textbf{b}) Top trace: ODMR on a~$\vert m_s=0\rangle$ to $\vert m_s=-1\rangle$ transition from several NV$^-$ centers inside a~hybrid ferro-diamond particle. Bottom trace: Time trace showing the~librational ring-down from the~hybrid structure. The~detection was realized using both the~laser scattered light and the~PL. Reprinted figure with permission from~\cite{Huillery2020}. Copyright 2020 by the~American Physical~Society.}\label{PL_motion}
\end{figure}

{In order for NV centers to efficiently detect the~center of mass motion, \mbox{Gieseler et al. ~\cite{Gieseler2020a}}  used broadband magnetic noise to excite the~motion of the~trapped micro-magnet (diameter $\approx 15~\upmu$m).  The magnetic field thermal noise at the~NV location, generated by the~100 $~\upmu$m distant driven trapped magnet, was then observed in~the~power spectral density (PSD) of the~NV PL evolution. Thanks to the~high quality factor of the~oscillator ($Q\approx 10^6$), tuning a~microwave to the~blue side of the~ODMR signal, as shown in~the~top of Figure~\ref{PL_motion}a, resulted in~a~sharp peak at the~mechanical oscillator frequency. }

Similarly, Huillery et al.~\cite{Huillery2020} reported NV-based detection of the~motion of a~$10~\upmu$m hybrid ferromagnet/diamond particle.
The latter was levitated in~a~Paul trap and the~libration was detected in~the~time domain after parametric excitation of the~magnetically confined librational mode. 
Figure~\ref{PL_motion}b shows the~ODMR (top trace) and the~ring down of the~librational mode (below) detected both using scattered light and the~NV PL. 
Notably, the~PL change was delayed with respect to the~instantaneous particle motion due to a~time lag between the~motion and the~magnetization. 

These studies open a~path towards the~NV read-out of the~Brownian motion of trapped harmonic oscillators.  
To reach this limit, more involved dynamical decoupling (DDC) techniques~\cite{Kolkowitz} could be used, with perspectives for attaining the~zero-point motion sensitivity of the~oscillator. Rabi oscillations, Ramsey and spin echoes from NV centers in~a~trapped diamond have in~fact been observed already without significant deterioration of the~$T_2$ from either the~charge noise or the~angular Brownian motion~\cite{Delord2018}.
Rabi oscillations were also observed in~\cite{Pettit2017} using an~optically trapped nanodiamond under weak magnetic fields.
DDC would, however, require to enter the~regime where the~trapping frequency of the~trapped mechanical oscillator exceeds the~decoherence rate $\Gamma_2^*$ of the~NV spins. We~discuss ways to achieve this in~Section~\ref{sec:production} .

\section{Magnetic Forces and Torques on a~Trapped Particle from the~Spin of NV$^-$ Centers}%
\label{sec:magnetic_force_torque}

We have discussed experiments where the~trapped particle motion was read-out by NV spin states. 
In order to strongly couple a~mechanical oscillator to spins, however, not only should the~mechanical motion affect the~spin state, but the~spin should also alter the~motion. 

In this section, we review work on the~reverse process where NV centers act on the~motion. We start by a~discussion on optimum sensing of the~force and the~torque induced by NV centers. 
 
\subsection{Force and Torque Sensitivity}

Presently, direct optical read-out of the~motion of trapped particles using scattered light is more efficient than using embedded NV centers.
Most experiments that use optical read-out are currently in~the~regime where the~influence of collisions with the~background gas dominates.
The detection noise is thus determined by the~resulting Brownian motion of the~levitating object. Note that the~ultimate sensitivity should be given by the~quantum back-action of the~measurement under low vacuum. This limited was recently reached in~experiments with trapped nano-spheres~\cite{Laan,Vijay}.
  
The most sensitive way to detect an~external torque, with current technology, is to use optical interferometric detection and to modulate the~torque at the~mechanical frequency of the~oscillator.
Modulation of the~amplitude of the~NV induced spin-torque can be done straightforwardly by modulating the~microwave tone that excites them. 
The minimum torque $\tau^{\rm min}_s$ that can be detected in~a~time $\delta T$ is then obtained by balancing 
the resulting signal amplitude and the~standard deviation of the~Brownian motion noise. One finds
\bea
\tau_s^{\rm min}\sqrt{\delta T}=\sqrt{4 k T \gamma_\theta I}=\sqrt{\frac{4 kT K_t}{Q \omega_\theta}},
\label{eq:torque_sensitivity}
\eea
where $Q=\omega_\theta/\gamma_\theta$ is the~quality factor of the~mechanical oscillator libration, and $K_t$ is the~trap rigidity, namely $I \omega_\theta^2$.

Tethered torque sensors with sensitivities in~the~$10^{-23}$ N$\cdot$m/$\sqrt{\rm Hz}$ range are realized nowadays with state-of-the-art nano-fabricated oscillators~\cite{Kim}. The~largest sensitivities with levitating systems have been achieved using levitating silica nanospheres that are attached to form a~dumbbell~\cite{Ahn}. The~authors reached a~record sensitivity of $10^{-28}$~N$\cdot$m/$\sqrt{\rm Hz}$. Trapped cristalline particles currently feature a~lower torque sensitivity than trapped amorphous particles, partly because of their currently lower quality factor. 
A sensitivity of $10^{-23}$ N$\cdot$m/$\sqrt{\rm Hz}$ was attained in~\cite{Huillery2020} using soft ferromagnetic particles at room temperature and at pressure levels in~the~$10^{-2}$ mbar range, very close to the~state-of-the-art torque sensing obtained at dilution fridge temperatures~\cite{Kim}.

A similar formula can be obtained for the~smallest detectable force acting on the~center of mass mode, by replacing 
$Q$ with $\omega_z/\gamma_z$, $K_t$ by $m \omega_z^2$ and $I$ by $m$.
Sensitivities in~the~zeptonewton$/\sqrt{\rm Hz}$ range were reported experimentally~\cite{Gieseler2012, Hempston, Ranjit1}. Very large sensitivities are also predicted for a~magnet levitating on top of a~punctured superconductor sheet in~the~Meissner state~\cite{PratCamps}. 
Amongst all spin--mechanical systems, trapped magnets currently feature record sensitivities in~the~$10^{-18}$N$/\sqrt{\rm Hz}$ range~\cite{Gieseler2020a, VinanteA}. Note that at the~low pressure levels employed in~these experiments ($\approx 10^{-5}$ mbar), damping is not determined by collisions with the~background gas so the~above model may not apply directly.

These last achievements not only show the~capabilities of levitating systems, close to the~sensing capabilities of MEMS, but also offer immediate perspectives for entering the~quantum regime by coupling magnets to distant NV centers.

\subsection{Observing NV Static Spin-Dependent Torque and Force}%

In this section, we discuss the~parameters required for observing spin-dependent forces and torques on a~trapped particle. 
We focus on the~expected static shifts, assuming that the~NV$^-$ centers have a~lifetime that is greater than the~typical time required to shift the~angle or the~center of mass. The~latter is typically on the~order of the~period of the~potential. 

\subsubsection{Angular Displacement Using NV$^-$ Centers}

Using Equation~\eqref{eq:Ham_lib'''}, one finds that the~torque applied to the~diamond in~the~magnetic state $\vert - 1'\rangle $ reads 
\begin{align}
\tau_s=-\left \langle \frac{\partial \hham'''_{\rm lib}}{\partial \theta} \right \rangle =-\hbar N \beta_{-1}.
\end{align}

The optimum shift will be found for $\theta=\pi/2$, where $\tau_s=-\hbar N \gamma_e B$ so the~largest single spin torque can be $\approx 10^{-25}$ N$\cdot$m.

\begin{figure}[!htbp]
 {\includegraphics[width=0.70\textwidth]{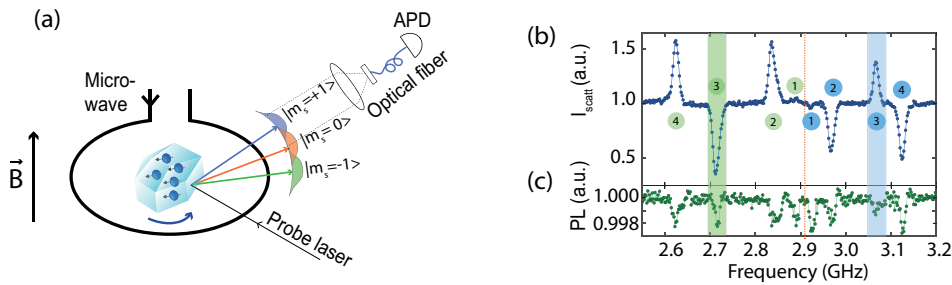}}
  \caption{{(\textbf{a}) Sketch} showing a~NV-doped diamond in~a~Paul trap where the~probe laser reflection angle from the~diamond surface depends on the~spin state. (\textbf{b}) Mechanically detected magnetic resonance observed by scanning the~microwave tone about the~NV transitions. (\textbf{c}) ODMR in~similar experimental conditions. Adapted from~\cite{Delord2020}.
 }\label{MDMRdelord}
\end{figure}

Let us consider a~particle with a~diameter of  15 $~\upmu$m undergoing Brownian motion in~a~harmonic trap under 0.1 mbar. Stable harmonic confinement is readily attainted at 300~K in~a~Paul trap with these parameters~\cite{Delord2020}, enabling a~sensitivity of $\approx 10^{-21}$~N$\cdot$m/$\sqrt{\rm Hz}$. This sensitivity is far from enabling single spin torque detection in~a~reasonable amount of time.
Reference~\cite{Delord2020} reported the~observation of spin-torque in~this parameter regime, albeit using $10^9$ spins that were all identically coupled with the~libration.
Figure~\ref{MDMRdelord}-{b}
 demonstrates mechanical detection of the~$\approx 10^{-19}$~N$\cdot$m spin torque averaged over a~few minutes with a~large signal to noise ratio (taken from~\cite{Delord2020}), revealing a~novel efficient method to probe ESR from NV~centers. 

In order to approach the~single spin torque level, particles with much smaller moment of inertia must be used. A~levitating particle with a~diameter of 100~nm and a~modest $Q$ factor of $10^4$ would be sufficient to reach sensitivities in~the~$10^{-25}$~N$\cdot$m/$\sqrt{\rm Hz}$ range. Single-spin torque could then be discerned within one second in~this experimentally achievable~regime.

\subsubsection{Center of Mass Displacement Using NV$^-$ Centers}

A similar discussion can be made to estimate the~magnitude of the~magnetic force on the~CoM of a~trapped diamond subjected to a~magnetic field gradient
\be
F_s = -\left \langle \frac{\partial \hham_{\rm com}}{\partial z} \right \rangle = -\hbar N \gamma_e \frac{\partial B_z}{\partial z }.\ee

Under a~gradient of $10^5$ T/m, one finds a~single-spin force of $10^{-19}$ N. 
These large gradients typically require trapped micron-size magnets a~few hundreds of nanometers from the~NV center. 
Although less attractive for quantum applications, ensembles of spins could be used to bypass this technical difficulty.

The spin-dependent force has not been observed thus far using NV$^-$ centers to the~best of our knowledge, even with large ensembles of spins. Even if strong magnetic field gradients can be achieved, it may induce a~large inhomogeneous broadening when using ensembles, which will forbid efficient microwave excitation of the~whole ensemble of NV$^-$ centers in~the~selected orientation.
The magnetic field gradient will thus be bounded by the~microwave excitation Fourier width $\delta \nu$ in~a~pulsed excitation scheme, so that  $\partial B_z/\partial z\vert_{\rm max}=\Delta \nu /\gamma_e d$, where $d$ is the~diameter of the~probed spin-ensemble.
Note that $\Delta \nu$ can reach more than 100 MHz using high power amplifiers and dedicated fast electronics~\cite{fuchs2009gigahertz}.

One further difficulty with NV centers coupled with trapped particles is to distinguish the~center of mass motion from spin-induced torques. Indeed, a~magnetic field offset is required in~order to magnetize the~NV$^-$ centers with a~microwave which will rotate the~trapped particle. 
The spin-induced angular displacements may thus contribute at the~same level as the~center of mass shifts, which complicates measurement analyses.
One solution is to align the~magnetic field along the $[111]$ direction of the~diamond, where no torque should be applied to the~particle.
This is a~notoriously difficult task when using levitating particles where the~angle between the~diamond crystalline axes and the~main trap axes cannot not always be controlled. 

Let us conclude by adding that single-spin forces were measured using Magnetic Resonance Force Microscopy (MRFM) in~\cite{Rugar}. Acquisition times of many hours and carefully engineered modulation techniques were employed. It is likely that significant progress can be made by applying these techniques to spin-mechanics with NV centers, with fascinating prospects for controlling the~motion of micro-objects using single long-lived qubits. 
\section{Dynamical Resonant Spin--Mechanical Interaction}%
\label{sec:dynamical}

Spin--mechanical systems can show richer physics than just static torques and forces. When calculating the~torque and force in~Sections~\ref{sec:sensing_motion} and \ref{sec:magnetic_force_torque},  it was implicitly assumed that the~spin population does not change after applying the~microwave excitation. This is not the~case in~the~strong spin--mechanical coupling regime. 
Indeed, after the~microwave excites the~spin and gives rise to a~torque (or to a~force), the~particle angle (or position) changes, which in~turn changes the~microwave detuning with respect to the~ESR, hence the~magnetization. We discuss the~resulting bistable and spin-spring effects below. 

Further, when the~spin-torque resulting from the~combined laser and microwave induced magnetization is lagging behind the~motion, exchange of heat between the~spin and the~mechanical oscillator can take place. Experiments have recently entered this regime~\cite{Delord2020}. 

We provide a~simplified analytical theory of such spin dynamical back-action. We~focus on the~angular degree of freedom, but the~calculation can be straightforwardly carried out for the~center of mass by making the~replacement $(\theta, G_\theta)\rightarrow (z, G_z)$. 

\subsection{Bistability and Spin-Spring Effect}

As in~opto-mechanics, bistability and modified spring constants can occur.  We first describe these two effects, in~the~limit where the~microwave magnetization and the~laser polarisation rates are faster than the~mechanical oscillator frequency. This means that no energy exchange between the~motion and the~spin can take place.

\subsubsection{Spin-Bistability}

The torque $\tau_s$ exerted on the~particle due to the~spin--mechanical coupling can be evaluated using Equation~\eqref{TLSHam}. We obtain 
\bea
\hat \tau_s=- \frac{\partial \hham''_{{\rm NV}+\mu\rm w }}{\partial \theta}=-\hbar N G_\theta \ket{1'}\bra{1'}.
\eea

In the~dispersive limit where $\Omega \ll \Delta(\theta)$, the~spins are mostly in~the~lowest energy eigenstate $\ket{-}$ of $\hham''_{{\rm NV}+\mu\rm w }$ where the~torque reads:
\bea
\tau_s = \bra{-} \hat \tau_s \ket{-}=-\hbar N G_\theta \vert \bra{-}\ket{1'} \vert^2 \approx - \hbar N G_\theta \left(\frac{\Omega}{\Delta(\theta)}\right)^2.
\eea

This spin torque can be added to the~restoring torque of the~levitating system. The~new angular stable position can be found by solving the~equation $\tau_{s} + \tau_{\rm trapping}=0$. It gives rise to a~third degree polynomial equation for $\theta$
\begin{align}
    \hbar N G_\theta \Omega^2 + I \omega_\theta^2 \theta (\Delta+G_\theta\theta)^2=0.
\end{align}

In the~very same way as in~opto-mechanics, two stable solutions for $\theta$ can be found when $\Delta < 0$. 
Angular bistability can then occur when the~microwave is swept across the~ESR transition. 

\subsubsection{Dynamical Backaction: Spin-Spring Effect}

{Linearizing about an~equilibrium position $\theta_0$, and introducing the~detuning \mbox{$\bar\Delta=\Delta-G_\theta\theta_0$,}} we get 
 \bea
\bra{-} \hat \tau_s \ket{-}\approx \tau_{s,0} +K_s (\theta-\theta_0),
 \eea
 where  
  \bea
 \tau_{s,0}= \hbar N G_\theta \left(\frac{\Omega}{\bar \Delta}\right)^2 ~~{\rm and}~~
 K_s= -2\hbar N G_\theta^2 \frac{\Omega^2}{\bar \Delta^3},
  \eea
 in~the~limit of small angle shifts. 
This expression predicts a~restoring torque in~the~limit where $\bar \Delta>0$ (blue detuned with respect to the~spin transition).

In the~presence of the~green laser light, transitions from the~dressed spin states will alter this predicted shift.  We will estimate it together with the~spin-cooling effect using a~density matrix formulation in~the~following. 

\subsection{Spin-Cooling}

In the~above estimation, we assumed that the~spins react immediately to a~change in~the~motion. When the~spin torque lags behind the~motion, friction forces can alter the~motional temperature and lead to spin-cooling, as calculated in~~\cite{Ge} and observed in~\cite{Delord2020}.
In order to evaluate the~dynamical back-action from the~spins to the~mechanical oscillator with retardation, one will include the~dissipation of the~electronic spin. 
Lastly, since most experiments are operating in~the~so-called adiabatic limit where the~frequency of the~mechanical oscillator is much smaller than the~spin dephasing rate, 
we will consider this regime for simplicity and discuss the~corresponding limits to spin-cooling. 

\subsubsection{Equations of Motion}

We will again assume that the~microwave is tuned to the~$ \ket{0'}$ to $ \ket{1'}$ transition.
When the~longitudinal decay time $T_1\approx$ ms is longer than the~time $1/\gamma_{\rm las} \ll 100~\mu$s it takes for laser polarizing the~NV spin, the~magnetic state $ \ket{-1'}$ is not populated. 
This can be ensured quite straightforwardly experimentally, using laser powers in~the~hundreds of $\upmu$W range with standard microscope objectives. We will assume that this is the~case here.
We can thus reduce the~study to the~two level system described in~Equation~\eqref{TLSHam}.

The von Neumann equation for the~reduced two-by-two spin density matrix $\hat\rho$ reads
\begin{align}\label{MB1}
\frac{\partial \rho_{10}}{\partial t} & = (-\Gamma_2^* + i \Delta(\theta) ) \rho_{10} +i \frac{\Omega}{2} (2 \rho_{11} -1)\\ 
\frac{\partial \rho_{11}}{\partial t} & = -\gamma_{\rm las} \rho_{11} + i \frac{\Omega}{2} (\rho_{10} - \rho_{10}^*), \label{MB2}
\end{align}
where $\Gamma_2^*$ is the~inhomogeneous broadening of the~NV$^-$ center and $\gamma_{\rm las}$ is the~optical pumping rate to the~$\ket{0'}$ state. We also assumed $\gamma_{\rm las}/2\ll \Gamma_2^*$, which is largely satisfied in~practice.
Note that these equations are valid when the~broadening is purely homogeneous and of a~Markovian nature. In~general, the~NV$^-$ centers couple to slowly fluctuating spin baths that generally implies Gaussian ESR lineshapes.

The above equations are coupled with the~equation of motion of the~particle via
\bea
I \frac{\partial^2 \theta}{\partial t^2} + I \gamma \frac{\partial \theta}{\partial t}+I {\omega_{\theta}}^2 \theta=\langle \hat \tau_s \rangle_\mathcal{B} + \tau_L,\label{MO}
\eea
where $\langle \hat \tau_s \rangle_\mathcal{B}=-\hbar N G_\theta \rho_{11}$ and $\mathcal{B}$ accounts for incoherent laser excitation to the~ground state, as well as pure dephasing due to dipolar coupling to the~P1 centers. 
Because of the~$\theta$ dependency in~$\Delta$, this system of equations is nonlinear. Here, we study the~system dynamics around a~steady-state, which will linearize the~set of equations. 

\subsubsection{Stationary Solutions}

We introduce the~steady-state quantities $\rho_{11}^0=\langle \rho_{11} \rangle$, $\rho_{10}^0=\langle \rho_{10} \rangle$, and $\theta_0=\langle \theta \rangle$, where $\langle .\rangle$ denotes time averaging. Writing the~incoherent pumping rate to the~magnetic state at the~angle $\theta_0$ as ${\Gamma_0}=\frac{\Omega^2\Gamma_2^*}{{\Gamma_2^*}^2+{\bar{\Delta}}^2}$, we get:
\bea
\rho_{11}^0=\frac{1}{2}\frac{\Gamma_0}{\gamma_{\rm las}+\Gamma_0}.
\eea

Using Equation~\eqref{MO}, one finds the~steady state solution for the~angle to be
\bea
I {\omega_{\theta}}^2 \theta_0=\langle \hat \tau_s \rangle^0_\mathcal{B}=-\hbar N G_\theta \rho_{11}^0.
\eea

This last equation gives a~third degree polynomial equation for $\theta_0$.
Depending on the~microwave detuning, there can either be one or two stable solutions for $\theta_0$. 

\subsubsection{Effective Susceptibility}

Writing each spin and angle parameters in~Equation~\eqref{MB1}--\eqref{MO} as $f(t)= f^0+\delta f (t)$ and transforming them to the~Fourier domain, these equations can be recast into the~compact expression $\delta \theta(\omega)=\chi_{\rm eff}(\omega)\delta\tau_L(\omega),$
where the~susceptibility $\chi_{\rm eff}(\omega)$ reads
\bea
\chi_{\rm eff}(\omega)&=&\frac{1}{I(\omega_\theta^2-\omega^2-i\omega \gamma)-K_s(\omega)}.
\eea

The quantity $K_s(\omega)$ is a~dynamical spin-rigidity which quantifies the~response of the~particle angle to a~change in~the~spin-torque. The~real part of $K_s(\omega)$ gives rise to a~shift of the~mechanical oscillator frequency while the~imaginary part gives rise to a~damping of the~mechanical motion.
We can rewrite the~susceptibility in~the~more condensed form 
\bea
\chi_{\rm eff}(\omega)&=&\frac{1}{I(\tilde\omega_\theta^2-\omega^2-i\omega \tilde\gamma)},
\eea
with 
\bea
\tilde\omega_\theta=\omega_\theta\big[1-\frac{\Re(K_s(\omega_\theta))}{2 K_t} \big] \quad {\rm and} \quad \tilde\gamma=\gamma\big[1+Q\frac{\Im(K_s(\omega_\theta))}{K_t}\big].
\label{eq:theory}
\eea

The modified damping and frequency of the~mechanical oscillator have been estimated in~the~limit $K_t \gg \Re(K_s(\omega_\theta))$, where $K_t=I\omega_\theta^2$ is the~trap rigidity. $Q=\omega_\theta/\gamma$ is the~quality factor of the~trapped particle.

\subsubsection{Dynamical Spin-Rigidity in~the~Adiabatic Limit}
%

In the~adiabatic limit $\vert \partial \rho_{10}/\partial t\vert  \ll \vert (-\Gamma_2^*+ i \Delta(\theta))  \rho_{10}\vert $, we have 
\bea
\frac{\partial \rho_{11}}{\partial t}= -\gamma_{\rm las} \rho_{11} -\frac{\Omega^2}{2\Gamma_2^*}\mathcal{L}(\theta)(2\rho_{11}-1), ~~{\rm where}~~
\mathcal{L}(\theta)=\frac{1}{1+(\Delta(\theta)/\Gamma_2^*)^2}.
 \eea
 
We now linearize this equation around the~steady states and move to the~Fourier space. We find:
\bea
K_s(\omega)= \hbar N \bar\Delta \frac{ (\alpha \tau)^2}{1+i \omega\tau},
~~{\rm where}~~\alpha=G_\theta\sqrt{\frac{ \gamma_{\rm las} \Gamma_0^2}{{\Gamma_2^*}\Omega^2}}~~{\rm and}\quad \tau=(\gamma_{\rm las}+\Gamma_0)^{-1}.
\eea

Finally, using Equation~\eqref{eq:theory}, we obtain
 \bea
\tilde\omega_\theta=\omega_\theta  \big[1+\frac{\hbar N }{2 K_t} \frac{ (\alpha \tau)^2}{1+(\omega_\theta\tau)^2}\bar\Delta \big]
~~{\rm and}~~
\tilde\gamma=\gamma\big[1-Q\frac{\hbar N  (\alpha \tau)^2 (\omega_\theta\tau)}{K_t  (1+(\omega_\theta\tau)^2)}\bar\Delta\big].
\label{eq:damping}
\eea

\begin{figure}[htbp!]
 {\includegraphics[width=0.70\textwidth]{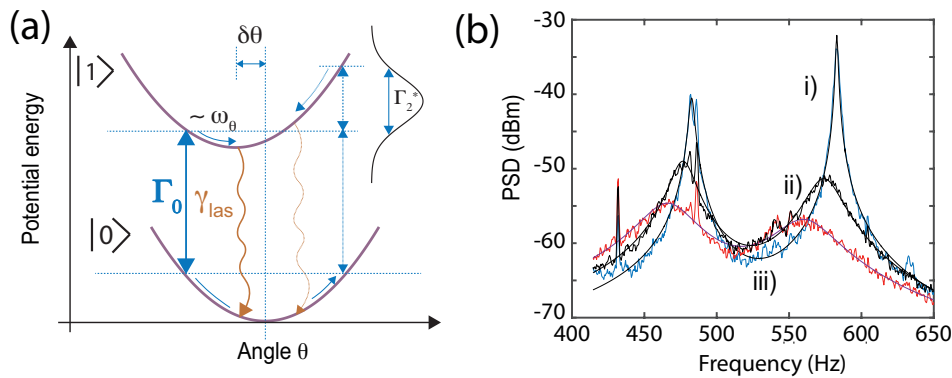}}
  \caption{(\textbf{a}) Spin-cooling 
 mechanism described in~the~adiabatic limit. The~two angular potential wells in~the~spin states $ \vert 0 \rangle$ and $\vert 1\rangle$ are offset from one another by $\delta \theta$ due to the~spin-torque. They are coupled via the~laser and microwave tone at the~rates $\gamma_{\rm las}$ and $\Gamma$, respectively, see text for explanations. Here the~microwave is tuned to the~red, enabling cooling. (\textbf{b}) PSD of two librational modes when the~microwave is tuned to the~blue (trace i) to resonance (trace ii) and to the~red (trace iii) of the~spin-resonance, respectively. Adapted from~\cite{Delord2020}.
 }\label{SC}
\end{figure}

As manifest in~Equation~\eqref{eq:damping}, when $\omega_\theta\tau$ is close to $1$ and $\overline \Delta <0$, the~oscillator motion can be damped, and thus cooled down. 
As in~cavity opto-mechanics, the~cooling is due to the~retarded nature of the~torque / force. 
Here, the~delay comes from the~finite time $\tau$ it takes to re-polarize the~spins in~the~magnetic state after the~oscillator is displaced from equilibrium, as depicted in~Figure~\ref{SC}a. 

The change in~damping induces a~change in~the~oscillator temperature. 
Using \linebreak $\delta \theta(\omega)=\chi_{\rm eff}(\omega)\delta\tau_L(\omega),$ one finds that the~variance of the~angle is
\bea
S_\theta(\omega)=\vert \chi_{\rm eff}(\omega)\vert^2  S_T,
\eea
where $S_T=2kT I \gamma$ is the~two-frequency correlation of the~Langevin torques. It is estimated at the~temperature $T$ of the~gas molecules surrounding the~particle, in~the~regime where $\hbar \omega_\theta \ll kT $. 
Using the~fluctuation dissipation theorem, one obtains a~simple relation between the~temperature $T_f$ in~the~presence of the~NV$^-$ spins and the~temperature $T$ of the~gas. One indeed has
$T_f=\frac{\gamma}{\tilde \gamma}T$.

For large quality factors and negative detunings, that is with a~microwave tuned to the~red, a~pronounced cooling can take place.
This effect is strongly analogous to Raman cooling or cavity cooling and was observed in~\cite{Delord2020} using diamonds levitated in~a~Paul trap.
Figure~\ref{SC}b shows the~power-spectral density of two libration modes of a~trapped diamond undergoing spin-cooling and spin-heating.
The limit to the~cooling efficiency will ultimately be given by back-action noise from either the~radiation pressure from the~laser beam that is used to polarise the~spins or by the~atomic spin noise. As in~cavity opto-mechanics, the~latter can be mitigated in~the~sideband resolved regime (SRR), where the~trapping frequency is larger than the~spin dephasing rate $\Gamma_2^*$. 

The SRR regime has not been attained so far using trapped particles coupled with NV$^-$ centers in~any platform, to the~best of our knowledge.
Several solutions are envisioned to increase the~trapping frequency and/or reduce the~spin linewidth. One of them consists of~coupling the~NV electronic spins to the~nuclear spins of the~nitrogen atoms~\cite{HuilleryCPT} to make use of the~very long nuclear spin lifetime. Another solution is to employ distant coupling schemes between NV centers in~ultra-pure diamonds and strongly confined ferromagnetic particles~\cite{Huillery2020}. If all other heat sources occur at a~rate smaller than the~spin-cooling in~the~SRR, ground state spin-cooling of the~libration or the~center of mass motion can then become a~reality, offering great perspectives for controlling the~quantum state of the~motion with the~NV spin. 

\section{Challenges Ahead for Levitated Spin-Mechanics}%
\label{sec:challenges}

Levitated spin-mechanics offers prospects for a~wide range of applications in~quantum science. The~negatively charge NV$^-$ center in~diamond is a~system of choice because of its robustness, which partly explains its increasing use in~present quantum technologies.
Nevertheless, recent experimental implementations have raised some technical difficulties. First and foremost, there is poor levitation stability when decreasing the~gas pressure. 
To date, optical trapping of diamond has been reported only above mbar pressures~\cite{Frangeskou2018NJP}, while electrodynamics and diamagnetic traps operate in~the~ $10^{-3}$ to $10^{-4}$ mbar range~\cite{Conangla, Hsu2015}. To~go further and operate close to the~standard quantum limit for instance, the~collisions with the~background gas must be suppressed. 
Furthermore, the~control of diamond in~terms of shape, size, and properties is not as straightforward as for silica, for example, where chemical processes allow the~production of mono-disperse spherical particles.
In the~present section, we highlight some of the~issues that can be tackled in~the~next generation of levitated spin--mechanical systems.

\subsection{Production of Diamond}%
\label{sec:production}
The diamonds that are employed for research purposes are almost always of artificial origin. They are produced mainly using three methods, detailed in~the~\mbox{Appendix~\ref{growth}}. The~most promising technique for achieving flawless diamonds is chemical vapour deposition (CVD). 
In the~different growth methods, the~NV concentration is controlled using different strategies that depend on the~initial nitrogen content in~the~diamond.
When the~sample contains large nitrogen concentrations, it can be irradiated by electrons or alpha particles to create vacancies. Diamond annealing then allows vacancy migration until being stably associated with a~nitrogen atom to form a~NV$^-$ center. The~irradiation dose dictates the~concentration of NV$^-$ centers per nano-diamond. Typically, with this technique, a~100~nm diamond can be doped to contain one, up to thousands of NV$^-$ centers.  
It has been shown that a~similar result can also be obtained using laser irradiation, with the~benefit of controlling the~position (and the~number) of the~produced NV$^-$ centers~\cite{Chen2017NP}. When starting with pure CVD diamonds, nitrogen is first implanted and converted to NV$^-$ centers, with a~few percent yield, during an~annealing process.

\subsection{Control of Diamond Shape and Properties}
As discussed previously, thanks to the~development of CVD grown diamonds and implantation techniques, it is possible to finely control the~purity, and the~number of hosted NV$^-$ centers in~the~diamond.
Over the~last decades, nano and micro-fabrication of diamond has enabled the~realization of increasingly complex diamond structures~\cite{Rani2021M}. Pillars of a~controlled aspect-ratio can be nano-fabricated, and even more advanced structures are achievable using reactive ion etching technics. 
This is of particular interest for trapping diamonds with large librational frequencies~\cite{Delord2017}.    

The development of coated diamond initially intended for biology applications may also increase the~achievable control of the~properties of trapped diamond. A~typical example is silica embedded diamonds, which allows obtaining spherical particles, and that have been shown to favor the~NV$^-$ centers luminescence stability in~optically levitated nano-diamonds~\cite{Neukirch2015}.

Spin-mechanics with trapped objects will thus benefit from the~tremendous progress in~diamond material science and continuously improved knowledge about diamond and NV$^-$ centers.  
A potential drawback is that the~production quantity of diamonds may be limited. Besides, one could be interested in~levitating a~specific diamond particle with physical properties that have been well-characterized beforehand. 
In both cases, statistical trapping procedures that start from a~sprayed colloidal particle solution, and rely on random trapping events are not ideal. The~different recent approaches proposed for in situ trapping, using a~piezo shacking, a~laser impulsion~\cite{Kuhn2015NL,Bykov}, or trapping particles embedded in~a~polymer thin film~\cite{Hsu2015}, are very encouraging for the~developments of on-demand trapping. Coupling these technics with an~in situ characterization using a~confocal microscope could solve these issues.

\subsection{Increasing the~NV$^-$ Concentration}

We have seen that increasing the~number of NV$^-$ centers that couple to the~oscillator motion is a~viable route towards observing strong spin--mechanical effects. 
While the~optimal density of NV$^-$ centers is generally a~compromise between sensitivity and coherence time $T_2^*$, other more exotic effects start to appear when the~spin density reaches a~critical~point. 

For concentrations that are larger than $\sim$1~ppm (corresponding to a~mean distance between spins of $\approx$15~nm or a~dipolar coupling strength of $\approx$15~kHz), the~dipolar coupling amongst NV spins plays an~important role. One important effect is the~modification of the~spin lifetime $T_1$ through dipolar coupling with other short lived NV$^-$ \mbox{centers~\cite{Jarmola_temperature_2012,mrozek_longitudinal_2015,choi_depolarization_2017}}. This particular effect has been used with a~levitating diamond in~a~Paul trap to observe a~resonant change in~the~spins' magnetic susceptibility~\cite{pellet2021magnetic}.

Other collective effects between NV$^-$ centers include the~cooperative enhancement of the~NV$^-$ centers' dipole interaction, a~phenomenon similar to that of super-radiance described in~\cite{Bachelard,PANAT,Venkatesh}, and observed with a~levitating diamond in~an optical tweezer by \mbox{Juan et al.~\cite{Juan}}.

\subsection{Internal Temperature of Levitated Diamonds}
\label{sec:heating}
Numerous studies have reported an~increase in~levitated diamond internal temperature under vacuum conditions ~\cite{Delord2017APL,Hoang2016a,Pettit2017, RahmanB}. This heating is detrimental for practical reasons since it may lead to the~burning or melting of the~levitated particle. Internal temperature also induces extra quantum decoherence channels, which may prevent the~observation of macroscopic quantum effects and impact the~contrast of the~spin resonance.

\begin{figure}[htbp!]
      
        \includegraphics[width=\linewidth]{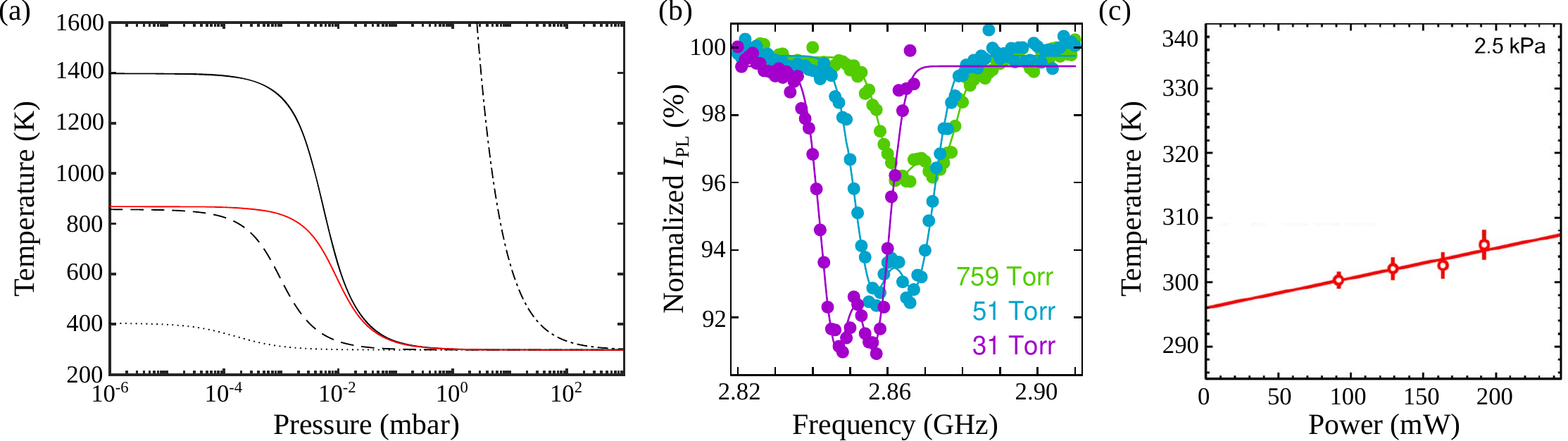}
        \caption{Internal 
 heating of levitated diamonds. (\textbf{a}) Expected internal temperature of optically levitated nano-diamonds of different purity as a~function of background gas pressure. The~temperature dependence is shown for standard commercial diamond (--$\cdot$--) up to the~best expected grade (dotted line). The~red-line corresponds to silica. Adapted from~\cite{Frangeskou2018NJP}. (\textbf{b}) Electron spin resonance from optically levitated nano-diamonds at different pressures. A~clear shift of the~central ESR frequency demonstrates diamond heating. Adapted from~\cite{Hoang2016a}. (\textbf{c}) Internal temperature of an~optically levitated nano-diamond hosting a~single NV$^-$ center as a~function of trapping laser power. Adapted with permission from~\cite{Pettit2017}. © The Optical Society.}%
        \label{fig:HeatingDiamond}
\end{figure}

This heating was shown to be induced by laser absorption by the~particle. The~final temperature is the~result of a~competition between absorption, heat conduction to the~surrounding residual gas, and black-body radiative exchange.  
Figure~\ref{fig:HeatingDiamond}a shows the~expected internal temperature for different diamond materials using the~model detailed in~references~\cite{Frangeskou2018NJP,Chang2010}. It can be seen that the~expected final temperature can vary by one order of magnitude depending on the~purity of the~diamond material.

Studies conducted on high purity CVD bulk diamonds have shown that a~very low laser absorbance is achievable~\cite{Webster2015JOSABJ}, even below the~silica absorption levels. Currently however, most experiments with levitated diamonds observe orders of magnitude higher heating rates. 
Any defects from the~diamond matrix may indeed worsen the~absorption 
First, NV$^-$ color centers may contribute to the~absorption due to non-radiative pathways of the~electron population (dashed lines in~Figure~\ref{fig:HeatingDiamond}a). 
Most importantly, other diamond matrix defects will also play their role, from isolated nitrogen atoms to graphite on the~diamond surface, through grain boundary or other atomic impurities~\cite{Walker1979RPPa}. 

The spin of the~NV$^-$ centers is actually an~invaluable tool to estimate the~diamond temperature ~\cite{Acosta2010b,Hoang2016a,Delord2017APL} because the~zero-field splitting $D$, between $m_s=0$ and $m_s=\pm 1$ states is temperature dependent, and well characterized ~\cite{Toyli2012}. One can measure the~internal temperature of a~levitated diamond by measuring the~ESR of the~NV$^-$ centers it hosts~\cite{Hoang2016a, Delord2017APL}, see Figure~\ref{fig:HeatingDiamond}b,c which shows heating measurements from trapped diamonds under vacuum.

Note that instead of improving the~properties of trapped diamonds that are destined to be trapped under a~magnetic field produced by a~fixed magnet, the~roles of the~diamond and the~magnet can be reversed. 
Schemes already discussed in~this review~\cite{Huillery2020, Gieseler2020a} involving trapped magnets were recently proposed. 
In these proposals, the~diamond containing NV$^-$ centers is attached to a~large heat sink, so it does not heat up significantly and can readily be made from CVD for improved spin-properties. Furthermore, no strong laser needs to be shone onto the~trapped magnet. 
\subsection{Beyond NV$^-$ Centers and Diamond}
The present review focused on NV$^-$ centers in~diamond, which is by far the~most studied system for levitated spin-mechanics. However, our discussions may be applied to other color centers with similar behavior, and also solve some of the~issues discussed in~this section. Typically, over recent decades, color centers in~SiC have been shown to own an~optically addressable spin resonance~\cite{Koehl2011N}. SiC can benefit from silicon-like technologies, which offer a~good level of control over the~material and its nano fabrication.
Ultimately, with the~recent isolation of single color centers~\cite{Redjem2020NE}, even silicon may become an~excellent platform for spin-mechanics.

\section{Conclusions}

In this review, we presented recent levitated spin-mechanics experiments, focusing specifically on NV$^-$ centers in~diamonds. We introduced a~formalism describing the~spin--mechanical interactions in~these experiments and highlighted the~advantages and limitations of this interaction scheme. 
We discussed the~technical challenges that remain towards taking the~full benefit of levitated spin-mechanics. The~common goals and emulation from atomic and solid-state physic, material science, and the~levitation scientific communities is a~cornerstone for success in~this field. 
This will undoubtedly lead to innovative and exciting experiments and applications.

\vspace{1cm}
\noindent This research was funded by the~\^Ile-de-France Region in~the~framework of DIM SIRTEQ and  by the Investissements d’Avenir of LabEx PALM (ANR-10-LABX-0039-PALM).\\
The authors would like to acknowledge fruitful discussions with Samuel Deléglise, Itsik Cohen, Cosimo Rusconi, Oriol Romero-Isart and Benjamin Stickler and thank James Millen for proof reading the~manuscript.

\noindent \textit{Authors Contributions~:}  Conceptualization, G.H., L.R. and M.P.; methodology, G.H., M.P., P.H. and L.R.; formal analysis, G.H. and M.P; writing---review and editing, G.H., L.R., C.P.-M., M.P. All authors have read and agreed to the~published version of the~manuscript.






\appendix

\section{Derivation of a~Simplified Hamiltonian}%
\label{Appendix A}
In this section, we detail the~calculation of the~simplified spin-libration Hamiltonian Equation~\eqref{TLSHam}. Several unitary transformations will be performed on the~Hamiltonian. For the~sake of clarity, we will divide our Hamiltonian into three different parts to do the~calculation and see the~effect of each transformation on the~different parts. We write 
\begin{align}
\hham= \hham_{\rm mecha} + \hham_{\rm NV}+\hham_{ \mu \rm w},
\end{align}
where
\begin{align}
\hham_{\rm mecha}=\frac{\hat{p}_{\theta}^2}{2I} +\frac{1}{2}I \omega_\theta^2 (\hat{\theta}-\theta')^2,
\end{align}
\begin{align}
\hham_{\rm NV}=\hbar D \hat{S}_{z'}^2 + \hbar \gamma_e B \hat{S}_{z},
\end{align}
and
\begin{align}
\hham_{\mu \rm w}=\hbar \Omega \cos({\omega t}) \hat{S}_{x}.
\end{align}

A few approximations have been made to derive this Hamiltonian. 
We suppose that $\gamma_e B$ is smaller than the~zero-field splitting $D$ of the~NV$^-$ center. We suppose the~transverse magnetic field $B_\perp$ and the~longitudinal magnetic field $B_\parallel$ to be of the~same order of magnitude. We also suppose that the~angular momentum of the~particle, given by the~mean value $\left \langle \hat{p}_\theta \right \rangle$, is a~few order of magnitude larger than the~typical spin momenta $\hbar N \left \langle \hat{S}_y \right \rangle$. This assumption is valid for micron-sized particles. Finally, we consider that $\Omega \ll \gamma_e B$.

\subsection{Moving to the~Particle Frame}

We move to the~particle frame by performing the~unitary transformation $\hat{U}=e^{i{\hat{\theta}} \hat{S}_y}$. This frame is relevant because the~eigenstates of the~new spin operator $\hat{S}_z$ are now the~ones where the~optical pumping process of the~green laser takes place.  In this frame, the~NV$^-$ Hamiltonian reads:
\begin{align}
\hham'_{\rm NV}= \hbar D \hat{S}_{z}^2 + \hbar \gamma_e B\left(\cos{{\hat{\theta}}} \hat{S}_{z}-\sin{{\hat{\theta}}} \hat{S}_{x}\right).
\end{align}

The mechanical part of the~Hamiltonian becomes
\begin{align}
\hham_{\rm mecha}'=\frac{(\hat{p}_{\theta}-\hbar \hat{S}_{y})^2}{2I} +\frac{1}{2}I \omega_\theta^2 (\hat{\theta}-\theta')^2.
\end{align}

One of the~assumption that we have made is to neglect the~spin contribution to the~total angular momenta of the~system, which means $\hbar N \left \langle \hat{S}_y \right \rangle \ll \left \langle \hat{p}_\theta \right \rangle$. Thus, we have: \begin{align}
\hham_{\rm mecha}' \simeq \frac{\hat{p}_{\theta}^2}{2I} +\frac{1}{2}I \omega_\theta^2 (\hat{\theta}-\theta')^2.
\end{align}

The microwave part of the~Hamiltonian simply becomes
\begin{align}
\hham_{\mu \rm w}'=\hbar \Omega \cos({\omega t}) \left(\cos{{\hat{\theta}}} \hat{S}_{x}+\sin{{\hat{\theta}}} \hat{S}_{z}\right).
\end{align}

\subsection{Diagonalization of the~NV$^-$ Hamiltonian}

The NV$^-$ part of the~Hamiltonian is diagonalized in~the~perturbative limit $\frac{\gamma_eB}{D} \ll 1$. We introduce the~operators  $\hat{u}_\perp=\frac{\gamma_eB}{D}\sin{\hat{\theta}}$ and $\hat{u}_\parallel=\frac{\gamma_eB}{D}\cos{\hat{\theta}}$. Thus, we have $\left \langle \hat{u}_\perp \right \rangle \ll 1$ and $\left \langle \hat{u}_\parallel \right \rangle \ll 1$. Using these operators, the~Hamiltonian of the~NV reads:
\begin{align}
\hham'_{\rm NV}= \hbar  D \hat{S}_{z}^2 + \hbar  D( \hat{u}_\parallel \hat{S}_{z}- \hat{u}_\perp \hat{S}_{x}).
\end{align}

We can treat the~second part of this Hamiltonian as a~perturbation and move to the~basis where this Hamiltonian is diagonal to second order in~$\frac{\gamma_eB}{D}$.
We consider the~unitary transformation:
\begin{align}
\hat{U}''=\begin{pmatrix} 1-\frac{\hat{u}_\perp^2}{4} &  \frac{\hat{u}_\perp}{\sqrt{2}} (1-\hat{u}_\parallel) & -\frac{\hat{u}_\perp^2}{4}  \\ -\frac{\hat{u}_\perp}{\sqrt{2}} (1-\hat{u}_\parallel)  &  1-\frac{\hat{u}_\perp^2}{2} & -\frac{\hat{u}_\perp}{\sqrt{2}} (1+\hat{u}_\parallel)\\ -\frac{\hat{u}_\perp^2}{4}  & \frac{\hat{u}_\perp}{\sqrt{2}} (1+\hat{u}_\parallel) & 1-\frac{\hat{u}_\perp^2}{4} \end{pmatrix}.
\end{align}

The new states we are considering by applying this transformation are defined by 
\begin{align}
\ket{+1'}&=\left(1-\frac{\hat{u}_\perp^2}{4}\right)\ket{+1}- \frac{\hat{u}_\perp}{\sqrt{2}} (1-\hat{u}_\parallel)\ket{0}-\frac{\hat{u}_\perp^2}{4}\ket{-1}\\
\ket{0'}&= \frac{\hat{u}_\perp}{\sqrt{2}} (1-\hat{u}_\parallel)\ket{+1}+\left(1-\frac{\hat{u}_\perp^2}{2}\right)\ket{0}+\frac{\hat{u}_\perp}{\sqrt{2}} (1+\hat{u}_\parallel) \ket{-1}\\
\ket{-1'}&=-\frac{\hat{u}_\perp^2}{4}\ket{+1}-\frac{\hat{u}_\perp}{\sqrt{2}} (1+\hat{u}_\parallel)\ket{0}+\left(1-\frac{\hat{u}_\perp^2}{4}\right)\ket{-1}.
\end{align}

We have $\hat{U}''^\dagger\hat{U}''=\hat{I}d + o(||(\hat{u}_\perp,\hat{u}_\parallel)||^2)$ which means that $\hat{U}''$ is unitary up to second order in~$||(\hat{u}_\perp,\hat{u}_\parallel)||$.
Under this transformation, the~NV Hamiltonian reads
\begin{align}
\hham''_{\rm NV} \simeq \hbar  D \begin{pmatrix} 1+\hat{u}_\parallel+\frac{\hat{u}_\perp^2}{2} & 0 & \frac{\hat{u}_\perp^2}{2} \\ 0  & -\hat{u}_\perp^2  & 0\\ \frac{\hat{u}_\perp^2}{2}  & 0 & 1-\hat{u}_\parallel+\frac{\hat{u}_\perp^2}{2}\end{pmatrix},
\end{align}
which can be written as:
\begingroup\makeatletter\def\f@size{8}\check@mathfonts
\def\maketag@@@#1{\hbox{\m@th\normalsize\normalfont#1}}%
\begin{align}
\hham''_{\rm NV} \simeq \hbar  D \begin{pmatrix} 1+\frac{\gamma_eB}{D}\cos{\hat{\theta}}+\left(\frac{\gamma_eB}{D} \right)^2 \frac{\sin{\hat{\theta}}^2}{2} & 0 & \left(\frac{\gamma_eB}{D} \right)^2 \frac{\sin{\hat{\theta}}^2}{2} \\ 0  & -\left(\frac{\gamma_eB}{D} \right)^2 \sin{\hat{\theta}}^2  & 0\\ \left(\frac{\gamma_eB}{D} \right)^2 \frac{\sin{\hat{\theta}}^2}{2} & 0 & 1-\frac{\gamma_eB}{D}\cos{\hat{\theta}}+\left(\frac{\gamma_eB}{D} \right)^2 \frac{\sin{\hat{\theta}}^2}{2} \end{pmatrix}.
\end{align}
\endgroup

Furthermore, we suppose that the~angle satisfies $\left \langle \hat{u}_\perp \right \rangle \simeq \left \langle \hat{u}_\parallel \right \rangle$ which implies that $\left(\frac{\gamma_eB}{D} \right)^2 \frac{\left \langle \sin{\hat{\theta}} \right \rangle^2}{2} \ll \frac{\gamma_eB}{D} \left \langle \cos{\hat{\theta}} \right \rangle $. We can then neglect non diagonal terms in~this regime and we obtain a~diagonal Hamiltonian:
\begingroup\makeatletter\def\f@size{8}\check@mathfonts
\def\maketag@@@#1{\hbox{\m@th\normalsize\normalfont#1}}%
\begin{align}
\hham''_{\rm NV} \simeq \hbar  D \begin{pmatrix} 1+\frac{\gamma_eB}{D}\cos{\hat{\theta}}+\left(\frac{\gamma_eB}{D} \right)^2 \frac{\sin{\hat{\theta}}^2}{2} & 0 & 0 \\ 0  & -\left(\frac{\gamma_eB}{D} \right)^2 \sin{\hat{\theta}}^2  & 0\\ 0 & 0 & 1-\frac{\gamma_eB}{D}\cos{\hat{\theta}}+\left(\frac{\gamma_eB}{D} \right)^2 \frac{\sin{\hat{\theta}}^2}{2} \end{pmatrix}.
\end{align}
\endgroup

Let us apply this transformation to the~mechanical part of the~Hamiltonian. The~unitary transformation $\hat{U}''$ only depends on $\hat{\theta}$ so it commutes with it. This transformation can be written $\hat{U}''=\hat{I}d+u\hat{V}(\hat{\theta})+o(|u|)$ with $u=\frac{\gamma_e B}{D}$ and $\hat{V}(\hat{\theta})=\hat{A}\sin{\hat{\theta}}$ where $\hat{A}=\frac{1}{\sqrt{2}}(\ket{+1}\bra{0}-\ket{0}\bra{+1}+\ket{-1}\bra{0}-\ket{0}\bra{-1})$. Furthermore, we have $\hat{V}(\hat{\theta})^\dagger=-\hat{V}(\hat{\theta})$. We~have:
\begin{align}
    \hat{U}''^\dagger\hat{p}_\theta \hat{U}''= \left(\hat{I}d-u\hat{V}(\hat{\theta})+o(|u|)\right)\hat{p}_\theta \left(\hat{I}d+u\hat{V}(\hat{\theta})+o(|u|)\right)
\end{align}
\begin{align}
    \hat{U}''^\dagger\hat{p}_\theta \hat{U}''= \hat{p}_\theta +u \commutator{\hat{p}_\theta}{\hat{V}(\hat{\theta})}+o(|u|) 
\end{align}
\begin{align}
    \hat{U}''^\dagger\hat{p}_\theta \hat{U}''= \hat{p}_\theta +u \hat{A} \commutator{\hat{p}_\theta}{\sin{\hat{\theta}}}+o(|u|) 
\end{align}
\begin{align}
    \hat{U}''^\dagger\hat{p}_\theta \hat{U}''= \hat{p}_\theta - i \hbar u \hat{A} \cos{\hat{\theta}}+o(|u|).
\end{align}

We can safely neglect the~second term under the~initial assumption $\left \langle \hat{p}_\theta \right \rangle \gg \hbar N u $. Thus, we~get:
\begin{align}
\hham_{\rm mecha}'' \simeq \frac{\hat{p}_{\theta}^2}{2I} +\frac{1}{2}I \omega_\theta^2 (\hat{\theta}-\theta')^2.
\end{align}

Furthermore, this transformation does not affect the~microwave Hamiltonian to first order in~$\frac{\gamma_e B}{D}$ so we get:
\begin{align}
\hham_{\mu \rm w}'' \simeq \hbar \Omega \cos({\omega t}) \left(\cos{{\hat{\theta}}} \hat{S}_{x}+\sin{{\hat{\theta}}} \hat{S}_{z}\right).
\end{align}

\subsection{Equilibrium Position of the~Paul Trap}%
\label{sec:energy}

We apply the~unitary transformation $\hat{U}'''=e^{i\theta' \hat{p}_\theta/\hbar}$ which redefines $\hat{\theta}$ as $\hat{\theta}-\theta'$. This transformation does not affect the~eigenstates of the~spin and we obtain to first order in~$\hat{\theta}$:
\begingroup\makeatletter\def\f@size{9}\check@mathfonts
\def\maketag@@@#1{\hbox{\m@th\normalsize\normalfont#1}}%
\begin{align}
\hham'''_{\rm NV} \simeq \hbar (\omega_{+1}+\beta_{+1}\hat{\theta}) \ket{+1'}\bra{+1'} + \hbar  (\omega_{0}+\beta_{0}\hat{\theta}) \ket{0'}\bra{0'}+\hbar (\omega_{-1}+\beta_{-1}\hat{\theta}) \ket{-1'}\bra{-1'},
\end{align}
\endgroup
with
\begin{align}
\omega_{+1}&=D+\gamma_eB\cos(\theta')+\frac{(\gamma_eB)^2}{D}  \frac{\sin{(\theta')}^2}{2}\\
\omega_{0}&=-\frac{(\gamma_eB)^2}{D} \sin{(\theta')}^2\\
\omega_{-1}&=D-\gamma_eB\cos(\theta')+\frac{(\gamma_eB)^2}{D}  \frac{\sin{(\theta')}^2}{2}\\
\beta_{i}&=\frac{\partial \omega_i}{\partial \theta'}.
\label{eq:energy}
\end{align}

This transformation shifts the~equilibrium position of the~mechanical oscillator by $\theta'$ and we get:
\begin{align}
\hham_{\rm mecha}''' \simeq \frac{\hat{p}_{\theta}^2}{2I} +\frac{1}{2}I \omega_\theta^2 \hat{\theta}^2.
\end{align}

The microwave Hamiltonian reads
\begin{align}
\hham_{\mu \rm w}''' \simeq \hbar \Omega \cos({\omega t}) \left(\cos({\theta'+\hat{\theta}}) \hat{S}_{x}+\sin({\theta'+\hat{\theta}}) \hat{S}_{z}\right).
\end{align}

\subsection{Rotating Frame of the~Micro-Wave}

The last unitary transformation is to move to the~microwave frame by performing the~unitary transformation $\hat U''''=e^{i\omega t \hat{S}_{z}^2}$. This transformation is diagonal so it commutes with the~NV$^-$ center Hamiltonian which is also diagonal. As it is a~time-dependent transformation, this will add a~shift in~energy to both the~$\ket{+1'}$ and $\ket{-1'}$ states which gives the~Hamiltonian:
\begin{eqnarray}
\hham''''_{\rm NV} &\simeq& \hbar  (-\Delta_{+1}+\beta_{+1}\hat{\theta}) \ket{+1'}\bra{+1'} + \hbar  (-\Delta_{0}+\beta_{0}\hat{\theta}) \ket{0'}\bra{0'}\\
&+&
\hbar  (-\Delta_{-1}+\beta_{-1}\hat{\theta}) \ket{-1'}\bra{-1'},
\end{eqnarray}
by defining $\Delta_{+1}=\omega-\omega_{+1}$, $\Delta_{-1}=\omega-\omega_{-1}$ and $\Delta_{0}=-\omega_{0}$.

The mechanical Hamiltonian is not affected by this transformation as it commutes with it.
Finally, under the~rotating wave approximation and by neglecting first order terms in~$\hat{\theta}$ which are negligible since $\Omega \ll \gamma_e B$, we obtain:
\begin{align}
\hham_{\mu \rm w}'''' \simeq \hbar \frac{\Omega}{2} \cos({\theta'}) \hat{S}_{x}.
\end{align}

Redefining $\Omega = \Omega \cos({\theta'})$, we get:
\begin{align}
\hham_{\mu \rm w}'''' \simeq \hbar \frac{\Omega}{2} \hat{S}_{x}.
\end{align}

Finally, the~Hamiltonian can be simply written as:
\begin{align}
\hham''''= \hham_{\rm mecha}'''' + \hham_{\rm NV}''''+\hham_{ \mu \rm w}'''',
\end{align}
where
\begin{align}
\hham_{\rm mecha}''''=\frac{\hat{p}_{\theta}^2}{2I} +\frac{1}{2}I \omega_\theta^2 \hat{\theta}^2.
\end{align}
\begingroup\makeatletter\def\f@size{8}\check@mathfonts
\def\maketag@@@#1{\hbox{\m@th\normalsize\normalfont#1}}%
\begin{align}
\hham''''_{\rm NV} \simeq \hbar  (-\Delta_{+1}+\beta_{+1}\hat{\theta}) \ket{+1'}\bra{+1'} + \hbar  (-\Delta_{0}+\beta_{0}\hat{\theta}) \ket{0'}\bra{0'}+\hbar  (-\Delta_{-1}+\beta_{-1}\hat{\theta}) \ket{-1'}\bra{-1'}
\end{align}
\endgroup
and
\begin{align}
\hham_{\mu \rm w}'''' \simeq \hbar \frac{\Omega}{2} \hat{S}_{x}.
\end{align}

\section{Diamond Synthesis}%
\label{growth}

There are three main diamond synthesis methods: 
\begin{itemize}
        \item The \emph{HPHT} process (high pressure, high temperature): A carbon precursor is brought under conditions of high pressure (typically > 5~GPa) and high temperature \linebreak (T~$\approx $~2000$~^\circ$C) in~order to create diamond. While this approach has been known since the~1950s, the~control of impurities in~the~diamond is not straigthforward. The~diamonds produced are often rich in~nitrogen impurities, typically around 200~ppm. Most recent works on diamond levitation used HPHT diamonds due to their ease of use and commercial availability. 
        \item The \emph{CVD} growth (chemical vapor deposition). A~reactor is used to deposit carbon atoms from a~methane gas, layer by layer on a~diamond substrate. It is then possible to finely control the~impurities present in~the~diamond. It is the~method of choice to create diamonds with very high purity. The~concentration of paramagnetic species such a~nitrogen or silicon can indeed be reduced below the~detection level, and the~concentration of $^{13}C$ atoms below natural abundance. Importantly, CVD growth also enables NV$^-$ center doping at any time during the~growth ~\cite{Achard2020JPDAP}.  
        \item \emph{Detonation} nanodiamonds are obtained by an~explosive reaction from a~carbon precursor.  This approach provides very small nanodiamonds, typically <10 nm, which are often highly graphitized~\cite{Bradac2010NN}. Such diamonds are thus not suited for the~applications discussed in~the~present review. 
\end{itemize}

Although recent efforts towards micro or nano-diamond creation have been made towards direct CVD~\cite{Feudis2020AMI,Tallaire2019AANM} and HPHT~\cite{Mindarava2020C} synthesis, the~general strategy is to mill bulk diamonds in~order to obtain nano (and even micro) diamonds. The~size of the~diamonds is then selected by centrifugation.


\end{document}